\documentclass[12pt]{article}
\pdfoutput=1
\usepackage{jheppub}
\usepackage{ytableau}
\usepackage[vcentermath]{youngtab}
\usepackage[utf8]{inputenc}

\newcommand\be{\begin{equation}}
\newcommand\ee{\end{equation}}

\newcommand\ch{\mathrm{ch}}
\newcommand\sh{\mathrm{sh}}
\newcommand\Tr{\mathrm{Tr}}

\title{Sphere correlation functions and Verma modules}
\abstract{We propose a universal IR formula for the protected three-sphere correlation functions of Higgs and Coulomb branch operators of
${\cal N}=4$ supersymmetric quantum field theories with massive, topologically trivial vacua.}
\author{Davide Gaiotto and Tadashi Okazaki}
\emailAdd{dgaiotto@perimeterinstitute.ca}
\emailAdd{tokazaki@perimeterinstitute.ca}
\affiliation{
Perimeter Institute for Theoretical Physics,\\
31 Caroline St. N., Waterloo, Ontario N2L 2Y5, Canada}
\begin{document}
\maketitle

%%%%%%%%%%%%%%%%%%%%%%%%%%%%%%%%%%%%%%%%%%%
%%%%%%%%%%%%%%%%%%%%%%%%%%%%%%%%%%%%%%%%%%%
\section{Introduction}
%%%%%%%%%%%%%%%%%%%%%%%%%%%%%%%%%%%%%%%%%%%
%%%%%%%%%%%%%%%%%%%%%%%%%%%%%%%%%%%%%%%%%%%

Supersymmetric partition functions and supersymmetric indices are powerful and well-developed tools to study supersymmetric quantum field theories,
their local operators and extended defects. They are often computable by generalizations of 
the supersymmetric localization techniques of \cite{Pestun:2007rz}.

These tools are typically available for theories endowed with a certain minimum amount of supersymmetry, 
depending on the specific setup, with special structures emerging in more supersymmetric situations. 

The main subject of this note is a collection of protected correlation functions of local operators on a three-sphere, which are available for 3d theories 
endowed with ${\cal N}=4$ supersymmetry. 

The three-dimensional sphere partition function \cite{Kapustin:2009kz} or its ellipsoid $S^3_b$ deformation \cite{Hama:2011ea} are well-defined for 
three-dimensional theories with ${\cal N}=2$ supersymmetry. They depend on the squashing parameter $b$ and on a collection of ``real masses'' 
associated to global symmetries. 

A three-dimensional theory with ${\cal N}=4$ supersymmetry can be treated as an ${\cal N}=2$ theory with a special global symmetry 
generator arising from the ${\cal N}=4$ R-symmetry. When the corresponding real mass is tuned to a particular value, the $b$-dependence drops out and 
the partition function acquires new features. Lacking a better name, we will refer to it as the ``special sphere partition function''. \footnote{See Appendix \ref{app:Sb} for details on this specialization. 
As discussed below, it is an analogue of the Schur index.}

The special sphere partition function can be enriched by a variety of BPS observables which are only available in theories with ${\cal N}=4$ supersymmetry.
The BPS observables of 3d ${\cal N}=4$ theories include local operators whose expectation values define the Higgs and Coulomb branches of the 
theory \cite{Dedushenko:2016jxl, Dedushenko:2017avn, Dedushenko:2018icp}. 

\begin{figure}
\begin{center}
\includegraphics[width=6.5cm]{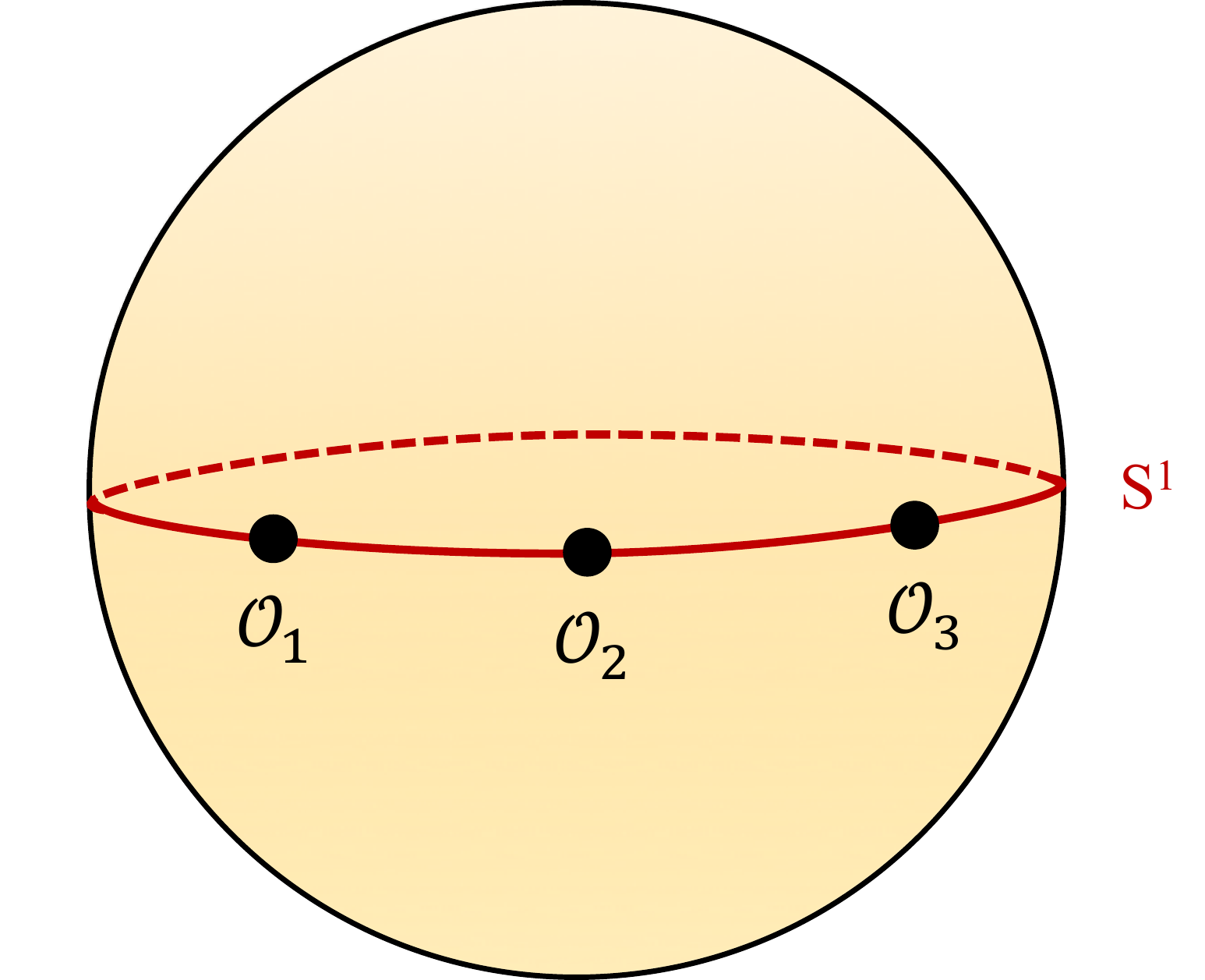}
\caption{
A collection of Higgs/Coulomb branch operators $\mathcal{O}_{i}$ along a great circle $S^1$ in the $S^3$. 
}
\label{figsphere}
\end{center}
\end{figure}

One can either decorate the special sphere partition function by a collection of Coulomb branch local operators at any point along a specific great circle $S^1$ in the $S^3$,
or by a collection of Higgs branch operators (see Figure \ref{figsphere}). 
These correlators will behave as a twisted trace on the quantized Coulomb or Higgs branch algebras of the 3d theory, as defined in \cite{Yagi:2014toa,Bullimore:2015lsa,Nakajima:2015txa,Braverman:2016wma}. 

The special sphere partition function $Z(m;\zeta)$ depends on two sets of parameters, ``masses'' $m$ and ``FI parameters'' $\zeta$. 
The quantum Coulomb branch algebra ${\cal A}^C_m$ depends on the masses and the twisting in the corresponding trace $\Tr^C_\zeta$ depends on the FI parameters. 
The opposite is true for the quantum Higgs branch algebra ${\cal A}^H_\zeta$. Overall, we must have relations
\begin{equation}
Z(m;\zeta) = \Tr^C_\zeta 1_C =  \Tr^H_m 1_H
\end{equation}
where $1_C$ and $1_H$ are the units in ${\cal A}^C_m$ and ${\cal A}^H_\zeta$ respectively. 

Several obvious questions arise. What selects $\Tr^C_\zeta$ and $\Tr^H_m$ among all possible twisted traces on the respective algebras?
Do they have special mathematical properties? The space of traces on the Higgs/Coulomb branch algebra appears, for example, in the ``quantum Hikita conjecture''
\cite{2018arXiv180709858K}. 

In the main body of the paper we find strong evidence of an ``IR formula'' for the sphere correlators, applicable to theories with massive, trivial, isolated vacua. 
The formula makes manifest the expansion of the sphere correlators into a natural basis in the space of twisted traces.

Massive trivial vacua $v$ of 3d ${\cal N}=4$ theories are associated to Verma modules $V_v^C$ and $V_v^H$ for the quantum Coulomb/Higgs algebras \cite{Bullimore:2016hdc}. The association 
depends on a choice of chamber in the space of FI parameters or masses. The same data also determines certain effective 
mixed Chern-Simons couplings $k_v$ which enter in the central charge $m \cdot k_v \cdot \zeta$ of the vacuum $v$ \cite{Bullimore:2016nji}. 
This data is the zero-th level piece of the Symplectic Duality correspondence \cite{2012arXiv1208.3863B}. 

One can obtain twisted traces $\Tr^{V^C_v}_\zeta$ and $\Tr^{V^H_v}_m$ simply by taking weighted traces over the Verma modules. 
\footnote{The twisted trace of quantized Higgs branch algebra which only admits generic masses has been also discussed in \cite{Dedushenko:2019mzv} by reducing the VOA characters. }
Although the modules are infinite-dimensional, the highest weight condition make the traces convergent in an appropriate chamber in the 
spaces of masses or FI parameters. Furthermore, one observes that 
\begin{align}
\Tr^{V^C_v}_\zeta 1_C &= e^{2 \pi i m \cdot k_v \cdot \zeta} f^C_v(\zeta), \cr
\Tr^{V^H_v}_m 1_H &= e^{2 \pi i m \cdot k_v \cdot \zeta} f^H_v(m).
\end{align}
We conjecture that
\begin{equation}\label{eq:Z}
Z(m;\zeta) = \sum_v \omega_v e^{2 \pi i m \cdot k_v \cdot \zeta} f^C_v(\zeta)f^H_v(m).
\end{equation}
Here $\omega_v$ is a fourth root of unity which we expect to encode background gravitational or R-symmetry Chern-Simons couplings.

Furthermore, we conjecture that 
\begin{equation}\label{eq:OC}
\langle \mathcal{O}_C \rangle \equiv  \Tr^C_\zeta \mathcal{O}_C= \sum_v \omega_v f^H_v(m)  \Tr^{V^C_v}_\zeta \mathcal{O}_C
\end{equation}
and 
\begin{equation}\label{eq:OH}
\langle \mathcal{O}_H \rangle \equiv \Tr^H_m \mathcal{O}_H= \sum_v \omega_v f^C_v(\zeta)  \Tr^{V^H_v}_m \mathcal{O}_H
\end{equation}
where $\mathcal{O}_{C}$ and $\mathcal{O}_{H}$ are the Coulomb and Higgs branch operators respectively. 

Our formula implies surprising constraints on the Verma module characters. 
The left hand side must be invariant under crossing walls in the spaces of FI and mass parameters,
but the sum on the right hand side is completely reorganized across walls. Furthermore, individual terms on the right hand side 
have poles as a function of $m$ and $\zeta$ which are absent on the left hand side, and thus must cancel in the sum.

\subsection{Generalizations and open questions}
This paper leaves several natural open problems. 
\begin{itemize}
\item It should be possible to justify our conjectures by a careful localization analysis of the special sphere partition function, 
perhaps focussing on vortex configurations along one great circle of the three-sphere analogous to these studied in \cite{Bullimore:2016hdc}.
\item Most aspects of our conjecture could be proven by a careful combinatorial analysis of the residue structure of the localization integrals. 
\item We have not explored the connection to the quantum Hikita conjecture and other aspects of the Symplectic Duality program. It would be interesting to do so. 
\item A natural generalization involves theories with vacua which are massive but topologically non-trivial, as it may happen if there is a discrete unbroken gauge symmetry. 
The category ``${\cal O}$'' of nice highest weight modules has a more complicated structure and our formula would require important modifications. 
\item A factorization of the ellipsoid partition function for ${\cal N}=2$ theories with massive vacua was observed before 
\cite{Pasquetti:2011fj, Beem:2012mb}. It involves a sum of products of certain ``holomorphic blocks'' depending on $\exp i \pi b^{\pm 2}$. 
It would be nice to verify if the two factorizations will match as the R-symmetry real mass is specialized. It should be the case, as both factorizations 
emerge from a sum-of-residues evaluation of the localization integral. 
\item The sphere partition function can be enriched by Wilson loops and vortex loops, exchanged by mirror symmetry \cite{Assel:2015oxa}. 
The local operators on line defects define generalizations of the Higgs or Coulomb branch algebras of physical and mathematical interest \cite{Dimofte:2019zzj,2019arXiv190504623W}. The decorated sphere partition function gives a trace on these algebras. 
Our conjectures can be extended accordingly in the presence of line defects. 
\item The partition function $Z(0;0)$ at the superconformal point is particularly interesting. The quantum Higgs and Coulomb branch algebras 
at the superconformal point have an independent cohomological definition which endows them with unexpected unitarity properties
\cite{Beem:2013sza,Chester:2014mea, Beem:2016cbd,Etingof:2019guc}. Individual terms in our IR formula diverge at the superconformal point,
complicating a direct comparison. 
\item The partition function $Z(0;0)$ is an important observable in holography.
In the large $N$ limit, the partition functions for 3d SCFTs with M-theory duals 
provide gauge theory derivation of the $N^{\frac32}$ growth of the numbers of degrees of freedom of $N$ M2-branes predicted in \cite{Klebanov:1996un}. 
The $N^{\frac32}$ behavior of the $S^3$ partition function was firstly discovered in \cite{Drukker:2010nc} 
for ABJM model whose M-theory dual is $AdS_{4}\times S^7/\mathbb{Z}_{k}$. 
It also shows up in a large class of 
3d $\mathcal{N}=3$ \cite{Herzog:2010hf, Santamaria:2010dm, Gulotta:2011vp, Crichigno:2012sk} 
and $\mathcal{N}=2$ \cite{Martelli:2011qj, Cheon:2011vi, Jafferis:2011zi, Gabella:2011sg, Amariti:2011jp, Amariti:2011uw, Amariti:2012tj, Jain:2019lqb, Amariti:2019pky} Chern-Simons matter theories 
whose M-theory duals are $AdS_{4}\times Y_{7}$ where $Y_{7}$ are Sasaki-Einstein manifolds 
in such a way that the $S^3$ free energy is proportional to $N^{\frac32}$ with the coefficient depending on the volume of $Y_{7}$. 
It would be interesting to give an holographic interpretation for our formula, perhaps by turning on the analogues of $\zeta$ and $m$ 
on the supergravity side to study the dual to $Z(m;\zeta)$.
\item Although our formula should be valid for any ${\cal N}=4$ SQFTs, we will only test it on ``standard'' gauge theory examples. It would be interesting to extend the analysis to 
${\cal N}=4$ Chern-Simons theories \cite{Gaiotto:2008sd,Hosomichi:2008jd,Hosomichi:2008jb,Bagger:2006sk,Gustavsson:2007vu,Aharony:2008ug,Aharony:2008gk}. 
\item The ``Schur index'' \cite{Romelsberger:2005eg,Gadde:2011ik,Gadde:2011uv} is a specialization of the supersymmetric index of four-dimensional supersymmetric quantum field theories
 \cite{Kinney:2005ej,Romelsberger:2005eg}, available for theories with ${\cal N}=2$ supersymmetry.
 The Schur index can be enriched by a variety of BPS observables which are only available in theories with ${\cal N}=2$ supersymmetry, such as BPS line defects 
wrapping the $S^1$ factor of the geometry and placed at any point along a specific great circle in the $S^3$ \cite{Dimofte:2011py,Cordova:2016uwk}. 
\footnote{The Schur index can be also decorated by inserting certain BPS local operators, the so-called Schur operators \cite{Pan:2019bor}. 
Also see \cite{Dedushenko:2019yiw} for the inclusion of surface defects. }
These ``Schur correlators'' are analogous to the 3d Coulomb correlators. 
\footnote{The relation between the quantized algebras of 3d $\mathcal{N}=4$ theories and chiral algebras of 4d $\mathcal{N}=2$ SCFTs have been also addressed in \cite{Dedushenko:2019mzv,Pan:2019shz}. }
Each BPS line defect $L$ maps to 
an element $\hat L$ in a certain ``quantum Coulomb branch'' algebra \cite{Gaiotto:2010be} and the Schur correlation functions behaves as a (twisted) trace for that algebra, i.e. 
\begin{equation}
\langle \mathcal{O}_1 \mathcal{O}_2 \rangle = \langle \omega(\mathcal{O}_2) \mathcal{O}_1\rangle \equiv \Tr_\omega \mathcal{O}_1 \mathcal{O}_2
\end{equation}
where $\omega$ is an automorphism of the algebra induced from an $U(1)_r$ symmetry rotation by $2 \pi$ and the operator ordering in the trace 
is given by the order along the great circle in the sphere. 
The quantum Coulomb branch algebra of 4d ${\cal N}=2$ theories is known mathematically as ``quantum K-theoretic Coulomb branch''
\cite{Nakajima:2015txa,Braverman:2016wma}. For theories of class ${\cal S}$, it is the quantization of a character variety \cite{Gaiotto:2010be}. 
The Schur correlation functions can be computed by localization in the UV or by a surprising IR formula based on Seiberg-Witten theory \cite{Cordova:2015nma,Cordova:2016uwk}. 
The trace property imposes non-trivial constraints on the Schur correlation functions, which are still poorly studied. 
Indeed, the very existence of a (twisted) trace on the quantum Coulomb branch algebra is a non-trivial mathematical statement. 
It would also be interesting to characterize the ``Schur'' trace within the space of possible (twisted) traces of the algebra.  
See Appendix \ref{app:sh} for further comments.
\end{itemize}

%%%%%%%%%%%%%%%%%%%%%%%%%%%%%%%%%%%%%%%%%%%
%%%%%%%%%%%%%%%%%%%%%%%%%%%%%%%%%%%%%%%%%%%
\section{3d $\mathcal{N}=4$ gauge theories}
%%%%%%%%%%%%%%%%%%%%%%%%%%%%%%%%%%%%%%%%%%%
%%%%%%%%%%%%%%%%%%%%%%%%%%%%%%%%%%%%%%%%%%%
In this section we briefly review standard 3d $\mathcal{N}=4$ gauge theories and present our conventions. 
The basic data entering the definition of a renormalizable 3d $\mathcal{N}=4$ gauge theory is 
\begin{enumerate}
\item A compact Lie group $G$ as a choice of gauge group 
\item A linear quaternionic representation \footnote{It is also called a symplectic representation. } $\mathcal{R}$ of $G$ as a choice of matter content 
\end{enumerate}

%%%%%%%%%%%%%%%%%%%%%%%%%%%%%%%%%%%%%%%%%%%
\subsection{Field content}
%%%%%%%%%%%%%%%%%%%%%%%%%%%%%%%%%%%%%%%%%%%
%fields and supermultiplets
The fields of the theory are collected into a vectormultiplet $\mathbb{V}$ transforming in the adjoint representation of $G$ 
and a hypermultiplet $\mathbb{H}$ transforming in the quaternionic representation $\mathcal{R}$. 

%VM
The vectormultiplet $\mathbb{V}$ contains  
a gauge connection $A_{\mu}$, gauginos $\lambda$, three real scalar fields $\vec{\phi}$ $=$ $(\phi_{1},\phi_{2},\phi_{3})$ and auxiliary fields $D$. 

% HM
The hypermultiplets $\mathbb{H}$ contain $4 N_{H}$ real scalar fields which parametrize $\mathbb{R}^{4N_{H}}$ with a hyperk\"{a}hler structure 
and spinors $\psi$ as their superpartners. 
The quaternionic representation $\mathcal{R}$ of $G$ is a representation with the canonical hyperk\"{a}hler structure 
in such a way that $G$ acts as a subgroup of the hyperk\"{a}hler isometry group $USp(N_{H})$ $=$ $U(2N_{H})\cap Sp(2N_{H},\mathbb{C})$ of $\mathbb{R}^{4N_{H}}$. 
We denote the hypermultiplet fields as complex elements $Z^\alpha$ in $\mathcal{R}$

We often consider the case where the quaternionic representation $\mathcal{R}$ is the sum of two conjugate complex representations: $\mathcal{R}$ $=$ $R\oplus R^{*}$ and split the hypermultiplet scalars into pairs of complex scalar fields $(X,Y)$ $=$ $(X^a,Y_a)_{a=1}^{N_{H}}$ $\in$ $R\oplus R^{*}$.
%\footnote{
%The representation $R\oplus R^{*}$ would admit simple weakly coupled $\mathcal{N}=(2,2)$ half-BPS boundary conditions \cite{Bullimore:2016nji}. 
%When $\mathcal{R}$ is pseudo-real, $X$ and $Y$ are combined into a single set of fields. }
%and denote by $(\psi,\widetilde{\psi})$ $\in$ $R\oplus R^{*}$ their fermionic partners. 

%%%%%%%%%%%%%%%%%%%%%%%%%%%%%%%%%%%%%%%%%%%
\subsection{Symmetries}
%%%%%%%%%%%%%%%%%%%%%%%%%%%%%%%%%%%%%%%%%%%
%R-symmetry
The theories have R-symmetry group $SU(2)_{C}\times SU(2)_{H}$ 
where the two factors respectively rotate vector and hypemultiplet scalar fields. 
In other words, they are isometries which rotate the complex structures on the two branches of vacua parameterized by vector and hypemultiplet scalar fields. 
The gauge field $A_{\mu}$, gauginos $\lambda$, three vector multiplet scalars $\vec{\phi}$ and auxiliary fields $D$ transform as 
$({\bf 1}, {\bf 1})$, $({\bf 2}, {\bf 2})$, $({\bf 3},{\bf 1})$ and $({\bf 1}, {\bf 3})$ 
while the hypermultiplet scalars $(Z,\bar Z)$ and spinors $\psi$ transform as 
$({\bf 1},{\bf 2})$, $({\bf1}, \overline{{\bf2}})$, $({\bf 2},{\bf 1})$ and $(\overline{{\bf 2}},{\bf 1})$ under the $SU(2)_{C}\times SU(2)_{H}$. 

The theories have two types of global symmetries.
%G_H
The global symmetry group $G_{H}$, which is called a Higgs branch global symmetry, or simply a flavour symmetry, 
is the residual symmetry that rotates the $N_{H}$ hypermultiplets. 
It is formally described as the normalizer $N_{USp(N_{H})}(\mathcal{R}(G))$ of the gauge group $G$ inside $USp(N_{H})$, modulo the action of the gauge group: 
\begin{align}
\label{G_H}
G_{H}&=N_{USp(N_{H})}(\mathcal{R}(G))/\mathcal{R}(G). 
\end{align}

%G_C
When the theory has a $U(1)$ factor in the gauge group $G$, it has a global symmetry group $G_{C}$, 
which is called a Coulomb branch global symmetry, or simply a topological symmetry. 
It rotates the periodic dual photons $\gamma$ defined by $d\gamma = *d A_{U(1)}$ for each Abelian factor in $G$. 
It is the Pontryagin dual of the Abelian part of $G$: 
\begin{align}
\label{G_C}
G_{C}&=\mathrm{Hom}(\pi_{1}(G),U(1))\simeq U(1)^{\textrm{$\#$ $U(1)$ factors in $G$}}
\end{align}
and is only carried by monopole operators. In the IR, it may be enhanced to a non-Abelian group whose maximal torus is (\ref{G_C}), or an even larger group. 
The enhanced symmetry mixes order and disorder operators of the theory. 

The global symmetry group $G_{C}\times G_{H}$ commutes with the R-symmetry. 

%%%%%%%%%%%%%%%%%%%%%%%%%%%%%%%%%%%%%%%%%%%
\subsection{Mass parameters and vacua}
%%%%%%%%%%%%%%%%%%%%%%%%%%%%%%%%%%%%%%%%%%%
The Lagrangian of 3d $\mathcal{N}=4$ gauge theory can be determined by the data $(G,\mathcal{R})$ and the three dimensionful parameters: 
\begin{enumerate}
\item A gauge coupling $g^2_{\mathrm{YM}}$ for each gauge factor. 
\item Three mass parameters $\vec{m}$
\item Three Fayet-Iliopoulos (FI) parameters $\vec{\zeta}$
\end{enumerate}
The gauge coupling does not enter the protected quantities we consider in this paper. 

Mass parameters $\vec{m}$ transform as a triplet of $SU(2)_{C}$ and take values in the Cartan subalgebra $\mathfrak{t}_{G_{H}}$ of flavour symmetry $G_{H}$. 
They are obtained as constant background expectation values of vector multiplet for flavour symmetry group $G_{H}$. The special sphere partition function 
will depend on a single mass parameter for each generator of $\mathfrak{t}_{G_{H}}$, which we denote simply as $m$

FI parameters $\vec{\zeta}$ transform as a triplet of $SU(2)_{H}$ and take values in the Cartan subalgebra $\mathfrak{t}_{G_{C}}$ of flavour symmetry $G_{C}$. 
They are obtained as constant background expectation values of scalar fields of twisted vector multiplet for topological symmetry group $G_{C}$. 
The special sphere partition function will depend on a single FI parameter for each generator of $\mathfrak{t}_{G_{C}}$, which we denote simply as $\zeta$

Turning on generic masses and FI parameters, the classical vacuum equations read
\begin{align}
\label{v_eq}
[\vec{\phi},\vec{\phi}]&=0,& 
(\vec{\phi}+\vec{m})\cdot (X,Y)&=0,& 
\vec{\mu}+\vec{\zeta}&=0
\end{align}
where $\vec{\mu}$ is the classical moment map for the action of gauge group $G$. 

Our main conjecture applies to theories which have isolated massive vacua upon turning on generic masses and FI parameters. 
In such a vacuum, a collection of hypermultiplet fields gains a non-zero vev in order to satisfy the $\vec{\mu}+\vec{\zeta}=0$ moment map constraints. 
The hypermultiplet vev forces some vectormultiplet scalars to also get a diagonal vev proportional to the masses, in order to satisfy the remaining equations. 
The resulting vevs spontaneously break the gauge symmetry, combining the gauge field and some hypers into massive gauge bosons.
The remaining hypermultiplets are made massive by the masses and vectormultiplet vevs. 

In a massive vacuum, the flavour moment maps $\mu_H$ receive vevs which are linear in the $\vec \zeta$ parameters. The $3 \times 3$ matrix of central charges 
\begin{equation}
\vec m \cdot k_v \cdot \vec \zeta \equiv \vec m \cdot \vec \mu_H
\end{equation}
and the $m \cdot k_v \cdot \zeta$ specialization will play an important role for us.

%%%%%%%%%%%%%%%%%%%%%%%%%%%%%%%%%%%%%%%%%%%
\subsection{The quantized Higgs branch algebra}
%%%%%%%%%%%%%%%%%%%%%%%%%%%%%%%%%%%%%%%%%%%
The Higgs branch operator insertions in the special sphere partition function are certain position-dependent linear combinations of gauge-invariant 
polynomials in the hypermultiplet fields. After the dust settles, they can be labelled by elements of the 
``quantized Higgs branch algebra'' ${\cal A}_H$, a quantum Hamiltonian reduction of the Weyl algebra with generators in $\mathcal{R}$.

The quantized Higgs branch also arises as the algebra of topological local operators in a certain $\Omega$ deformation of the gauge theory \cite{Yagi:2014toa}. 

Concretely, the Weyl algebra is generated by symbols $Z^{\alpha}$ with commutator 
\begin{equation}
[Z^{\alpha},Z^{\beta}] = \omega^{\alpha \beta}.
\end{equation}
Here $\omega^{\alpha \beta}$ is the symplectic form on $\mathcal{R}$.

The algebra is equipped by an $USp(N_{H})$ action with generators given by the quantum moment maps 
\begin{equation}
\mu^{\alpha\beta} = Z^{(\alpha} Z^{\beta)}.
\end{equation}

The quantum Higgs branch algebra is generated by gauge-invariant operators, i.e. polynomials in $Z$ which commute with the gauge moment maps 
\begin{equation}
\mu_G = T_G \cdot \mu + \zeta
\end{equation}
where $T_{G}$ are the gauge generators and $\zeta$ the ``quantum'' Fayet-Iliopoulos  parameters, living in the Abelian subalgebra of the gauge group. 
We have to quotient the gauge-invariant subalgebra by the ideal generated by the $\mu_G$.

The quantized Higgs branch algebra is equipped with quantum moment maps $\mu_{H}$ generating the flavour symmetry $G_{H}$.

\subsection{The quantized Coulomb branch algebra}
The quantized Coulomb branch also arises as the algebra of topological local operators in a certain $\Omega$ deformation of the gauge theory. 

The most basic Coulomb branch BPS insertions are labelled by gauge-invariant polynomials in an adjoint field $\varphi$. These commute with each other and 
equip the quantized Coulomb branch algebra ${\cal A}_C$ with the structure of an integrable system. 

General Coulomb branch operators, though, are disorder (monopole) operators. A proper definition of the quantized Coulomb branch algebra ${\cal A}_C$
requires some mathematical subtlety \cite{Nakajima:2015txa,Braverman:2016wma}. See also  \cite{Gomis:2011pf,Ito:2011ea,Kapustin:2006pk}.
The algebra can be given an ``Abelianized'' description as a subalgebra of a shift algebra \cite{Bullimore:2015lsa}, generated by certain rational functions in the eigenvalues $\varphi_i$ of $\varphi$ and 
some shift operators $v_{n_*}$ with 
\begin{equation}
v_{n_*} \varphi_i = (\varphi_i  \pm n_i) v_{n_*}.
\end{equation}
The $v_{n_*}$ satisfy further theory-dependent relations involving the $\varphi_i$ and the ``quantum'' mass parameters $m$. 

%%%%%%%%%%%%%%%%%%%%%%%%%%%%%%%%%%%%%%%%%%%
\subsection{The special sphere partition function}
%%%%%%%%%%%%%%%%%%%%%%%%%%%%%%%%%%%%%%%%%%%
The special sphere partition function is given by an integral over the Cartan subalgebra of the gauge algebra: 
\begin{equation}
\label{Int_S3}
Z(m;\zeta) = \frac{1}{|\mathcal{W}(G)|}  \int_{-\infty}^\infty dx e^{2 \pi i \zeta \cdot x} \frac{\prod_{\alpha} 2 \sinh \pi \alpha \cdot x  }{\prod_A 2\cosh \pi w_A \cdot(x,m)}
\end{equation}
where $x$ lives in the Cartan subalgebra of $G$ and $|\mathcal{W}(G)|$ is the order of Weyl group of $G$. 
The numerator is a contribution from the vectormultiplet, 
%Vandermonde-like measure 
involving the roots $\alpha$ of the gauge algebra 
and the denominator has a $\cosh$ factor for each hypermultiplet, with $w_A$ being the weight of the $A$-th hypermultiplet under the 
gauge and flavour Cartan.

A few remarks are in order
\begin{itemize}
\item The partition function is originally defined for real $\zeta$ and $m$ and a standard integration contour along the real axis. 
It is only well-defined if the theory has ``enough'' matter, so that the denominator makes the integral converge. From now on, we will assume this is the case.
\item Turning on an imaginary part for $\zeta$ decreases the rate of convergence. The integral expression is well-defined in a ``physical'' strip in the $\zeta$ plane. The analytic continuation beyond the strip is possible, but will have poles. 
\item Turning on an imaginary part for $m$ risks pinching the contour of integration between poles of the integrand. Again, the partition function will be well-defined in a physical strip in the $m$ plane, but will have poles when analytically continued beyond that. 
\end{itemize}

%On general grounds, poles of the sphere partition function should be expected to correspond to flat directions of the path integral \cite{Gaiotto:2012xa}. 
%Poles at fixed values of $m$ should correspond to flat directions where Higgs branch operators receive vevs, while poles at fixed values of $\zeta$ should correspond to flat directions where Coulomb branch operators receive vevs.

\begin{figure}
\begin{center}
\includegraphics[width=10.5cm]{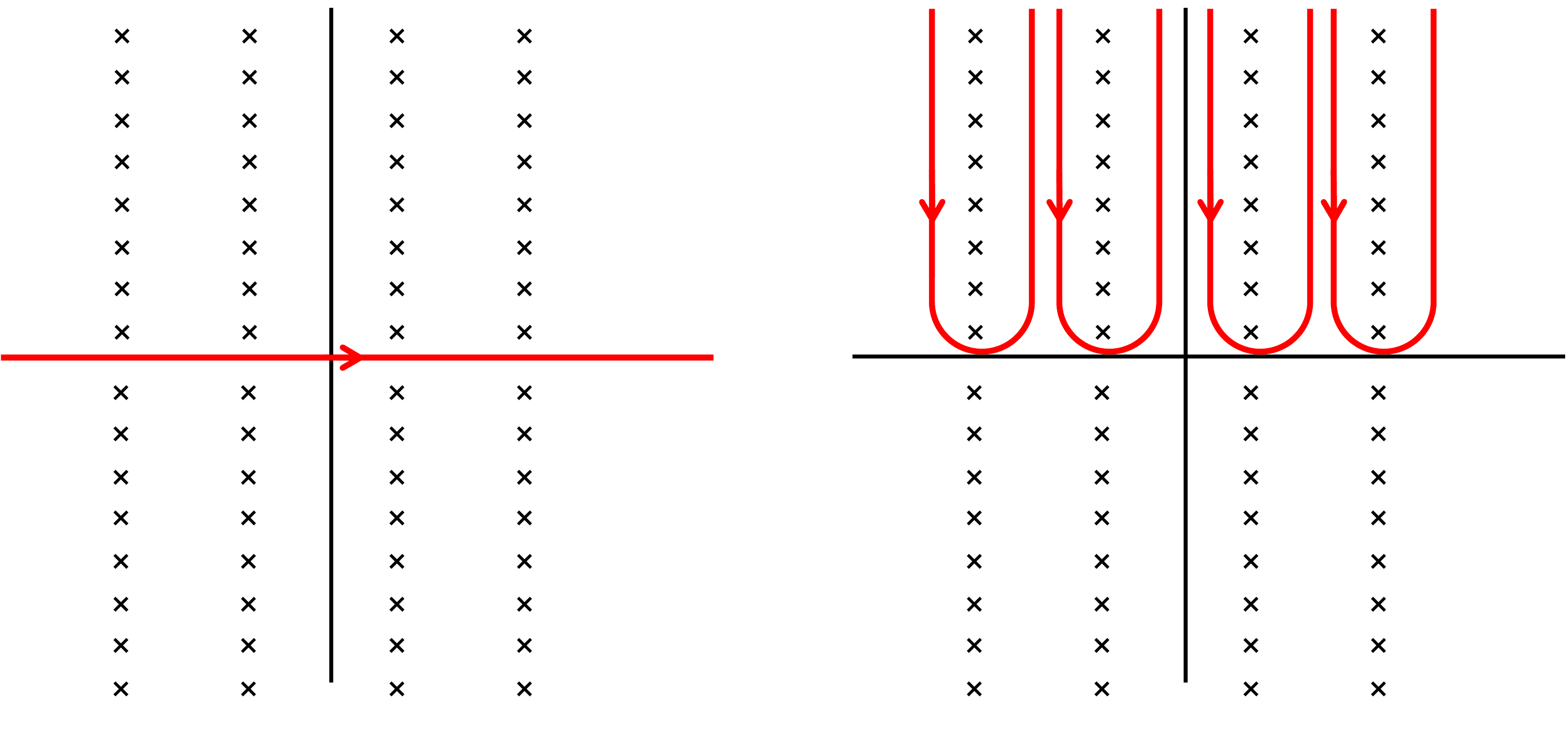}
\caption{
The integration contour in the complex $x$ plane along the real axis (left) is deformed for positive FI parameters $\zeta$ so that it encloses 
sequences of poles in the upper-half plane (right). 
}
\label{figpole}
\end{center}
\end{figure}

In favourable situations, picking a specific chamber for $\zeta$ allows one to close the integration contour at infinity and evaluate the integral as a sum over residues 
(Figure \ref{figpole}). 
We expect this to be the case for the massive theories we are interested in. As the integrand is periodic under $x_a \to x_a + i$, poles of the integrand will come in families 
whose residues differ by an overall power of $e^{2 \pi \zeta}$. Re-summing each family would already give an expression of the form 
\begin{equation}
Z(m;\zeta) = \sum_v \omega_v e^{2 \pi i m \cdot k_v \cdot \zeta} f^C_v(\zeta)f^H_v(m)
\end{equation}
for {\it some} choices of phases $\omega_v$, pairing $k_v$ and functions $f^C_v(\zeta)$ and $f^H_v(m)$. 

The $\cosh \pi w_A \cdot(x,m)$ factors which do not contribute to the pole, together with the Vandermonde factors, give the functions $f^H_v(m)$. 
The sum over residues in the family gives the functions $f^C_v(\zeta)$.

Associating each family to a choice of vacuum $v$ is a rather simple combinatorial problem. Each pole receives a contribution from a collection 
of $\cosh \pi w_A \cdot(x,m)$ factors, which we identify with the collection hypers which get a vev in the vacuum. The poles impose the same linear constraints 
on $x$ and $m$ which the vacuum vevs impose on $\vec \phi$ and $\vec m$. As a result, the $e^{2 \pi i \zeta \cdot x}$ factor evaluated at the residue matches the expected factor of 
$e^{2 \pi i m \cdot k_v \cdot \zeta}$ for the vacuum up to an overall power of $e^{2 \pi \zeta}$. 

The non-trivial statement is that this correspondence should be one-to-one for massive, topologically trivial vacua: we do not get multiple 
families from the same collection of $\cosh$ factors. We will see this combinatorics in detail in Abelian examples, but we will not attempt to find
a general proof.

The identification of the $f^C_v(\zeta)$ and $f^H_v(m)$ functions with Verma module characters is even less obvious. It is somewhat easier for
$f^C_v(\zeta)$, as the Coulomb branch Verma modules have a description in terms of vortex moduli spaces \cite{Bullimore:2016hdc}. 
 On the Higgs branch side, the characterization of the Verma modules for the quantum Hamiltonian reduction is a bit trickier. A physical discussion can be found in \cite{Bullimore:2016nji}.
Rather than trying to justify our conjecture for the Higgs branch Verma modules, we will focus on the Coulomb branch side and then appeal to mirror symmetry. 

The vortex construction labels a basis for the Coulomb branch module with equivariant fixed points on a vortex moduli space. Intuitively, each fixed point 
corresponds to some collection of hypermultiplets gaining a position-dependent vev with a zero of order $n_*$. We identify these equivariant fixed points 
with residues associated to the $n_a$-th zero of the $a$-th $\cosh$ factor. The value of $x$ at each residue matches the equivariant weight of the corresponding 
basis vector.

As a result, we expect to have an one-to-one correspondence between the non-zero residues in the contour integral and the equivariant fixed points on the vortex moduli space,
leading to the conjectural identification of $e^{2 \pi i m \cdot k_v \cdot \zeta} f^C_v(\zeta)$ with the character of the Verma module associated to the vacuum by the vortex construction.

%%%%%%%%%%%%%%%%%%%%%%%%%%%%%%%%%%%%%%%%%%%
\subsection{Higgs branch correlators}
%%%%%%%%%%%%%%%%%%%%%%%%%%%%%%%%%%%%%%%%%%%
As mentioned in the Introduction, the Higgs branch sphere correlators behave as twisted traces on 
${\cal A}_H$. The twist in the trace combines a $\pm 1$ sign from the action of the center of $SU(2)_H$ 
and a flavour rotation with parameter $2 \pi i m$. Concretely, 
\begin{equation}
\langle \mathcal{O}_{H_1} \mathcal{O}_{H_2} \rangle_{S^3} = (-1)^{F_1} e^{-2 \pi m_a q^a_1} \langle \mathcal{O}_{H_2} \mathcal{O}_{H_1} \rangle_{S^3}.
\end{equation}

They have a simple integral expression \footnote{See \cite{Dedushenko:2016jxl} for the derivation from localization. }
\begin{equation}
\langle \mathcal{O}_H \rangle_{m;\zeta} =  \int_{-\infty}^\infty dx e^{2 \pi i \zeta \cdot x} \left[ \prod_{\alpha} 2 \sinh \pi \alpha \cdot x\right]  \langle \mathcal{O}_H\rangle^{\mathrm{hyper}}_{m,x}
\end{equation}
where $\langle \mathcal{O}_H\rangle^{\mathrm{hyper}}_{m,x}$ is the correlation function in the free hypermultiplet theory, with 
mass parameters $m$ and $x$ for the flavour and gauge symmetry. 

As we review in the next section, the free correlation function $\langle \mathcal{O}_H\rangle^{\mathrm{hyper}}_{m,x}$ is fully determined by the twisted trace condition: 
the Weyl algebra has a canonical, unique twisted trace, naturally normalized so that 
\begin{equation}
\Tr^{\mathrm{hyper}}_{m,x} 1 = \frac{1}{\prod_A 2\cosh \pi w_A \cdot(x,m)}
\end{equation}
and computed by Wick contractions or as a trace on the unique highest weight module. 

Furthermore, if we promote $e^{-2 \pi x}$ to an element $g\in G^{\mathbb{C}}$ of the complexified gauge group, 
inserting a moment map $\mu$ in the twisted trace is the same as taking a derivative with respect to $g$. 

The $\sinh$-Vandermonde in the contour integral can be thought of as an analytic continuation of the usual Haar measure on $G$ to $G^{\mathbb{C}}$.
The whole integral can then be morally interpreted as an integral over a real cycle in $G^{\mathbb{C}}$. Integration by parts implements the 
quantum Hamiltonian constraint $\mu + \zeta=0$ in the correlation functions. 

This statement remains true no matter which integration contour we choose. That means taking a residue at any pole 
will still give a collection of correlation functions behaving as a twisted trace for ${\cal A}_H$. 
For a massive theory, this means we can expand 
\begin{equation}
\langle \mathcal{O}_H \rangle_{m;\zeta}= \sum_v \omega_v f^C_v(\zeta)  \Tr^{v}_m \mathcal{O}_H
\end{equation}
for {\it some} traces $\Tr^{v}_m$. 

Our conjecture (\ref{eq:OH}) requires the identification of these traces with actual traces over Verma modules. It is a trickier combinatorial problem, which we will 
not address here. The mirror statement is a bit simpler. 

\subsection{Coulomb branch correlators}
%%%%%%%%%%%%%%%%%%%%%%%%%%%%%%%%%%%%%%%%%%%
As mentioned in the Introduction, the Coulomb branch sphere correlators behave as twisted traces on 
${\cal A}_C$. The twist in the trace combines a $\pm 1$ sign from the action of the center of $SU(2)_C$ 
and a flavour rotation with parameter $2 \pi i \zeta$. Concretely, 
\begin{equation}
\langle \mathcal{O}_{C_1} \mathcal{O}_{C_2} \rangle_{S^3} = (-1)^{F_1} e^{-2 \pi \zeta_\alpha q^\alpha_1} \langle \mathcal{O}_{C_2} \mathcal{O}_{C_1} \rangle_{S^3}.
\end{equation}

An important simplification comes from the observation that the Abelianized expressions for the Coulomb branch 
generators are naturally compatible with the localization integral for the sphere correlators. Once we expand 
a given observable as a sum over shift operators
\begin{equation}
\mathcal{O}_C = \sum_{n_*} R_{n_*}(\varphi,m_{\mathbb{C}}) v_{n_*}
\end{equation}
we simply compute  \footnote{Also see \cite{Dedushenko:2016jxl,Dedushenko:2017avn,Dedushenko:2018icp}. }
\begin{equation}
\langle \mathcal{O}_C \rangle_{m;\zeta} =  \int_{-\infty}^\infty dx e^{2 \pi i \zeta \cdot x} \frac{\prod_{\alpha} 2 \sinh \pi \alpha \cdot x  }{\prod_A 2\cosh \pi w_A \cdot(x,m)} R_{0}(- i x,-i m).
\end{equation}

For a massive theory, the sum over residues at $x=x^*$ within each family would be modified by a factor of $R_{0}(- i x^*,-i m)$,
which is {\it not} periodic. Hence the insertion will modify the function  $f^C_v(\zeta)$.
The values of $x=x^*$ which appear in the sum can be recognized with the equivariant weights of the fixed points in 
the vortex moduli space which label states in the Verma modules in \cite{Bullimore:2016hdc}. This makes it very plausible that the 
sum over residues would compute the trace of $\mathcal{O}_C$ on the Verma modules, leading to (\ref{eq:OC}).

%%%%%%%%%%%%%%%%%%%%%%%%%%%%%%%%%%%%%%%%%%%
%%%%%%%%%%%%%%%%%%%%%%%%%%%%%%%%%%%%%%%%%%%
\section{Free hypermultiplets}
\label{sec_hyp}
%%%%%%%%%%%%%%%%%%%%%%%%%%%%%%%%%%%%%%%%%%%
%%%%%%%%%%%%%%%%%%%%%%%%%%%%%%%%%%%%%%%%%%%
It is instructive to look in detail at the twisted traces for the free hypermultiplet 
quantized Higgs algebra, i.e. the Weyl algebra generated by symbols $Z^{\alpha}$ with commutator 
\begin{equation}
[Z^{\alpha},Z^{\beta}] = \omega^{\alpha \beta}.
\end{equation}

The Weyl algebra has no untwisted trace: every element, including $1$, is a commutator and thus must have zero trace. 
On the other hand, it has interesting twisted traces. For example, consider the zero mass correlation functions, which satisfy 
\begin{equation}
\langle Z^\alpha \mathcal{O} \rangle =- \langle  \mathcal{O} Z^\alpha\rangle. 
\end{equation}

We can massage that to 
\begin{equation}
\langle Z^\alpha \mathcal{O} \rangle =\frac12 \langle [ Z^\alpha,\mathcal{O}]\rangle 
\end{equation}
which fixes all correlators as a sum over Wick contractions. 

If we turn on a generic twist, 
\begin{equation}
\langle Z^\alpha \mathcal{O} \rangle_M =- M^\alpha_\beta \langle  \mathcal{O} Z^\beta\rangle_{M} 
\end{equation}
with $M$ being a symplectic transformation, we get instead  
\begin{equation}
\langle Z^\alpha \mathcal{O} \rangle_{M} =\left[(1+M^{-1})^{-1}\right]^\alpha_\beta \langle [ Z^\beta,\mathcal{O}]\rangle_{M}. 
\end{equation}

\subsection{One hypermultiplet with $U(1)$ mass}
The most important example for us is a single hypermultiplet with an $U(1)$ mass parameter. 
Now we have 
\begin{equation}
[X,Y]=1
\end{equation}
with 
\begin{equation}
\langle X \mathcal{O} \rangle =- e^{2 \pi m} \langle  \mathcal{O} X\rangle,  \qquad \qquad  \langle Y \mathcal{O} \rangle =- e^{-2 \pi m} \langle  \mathcal{O} Y\rangle 
\end{equation}
so that 
\begin{equation}
\langle X \mathcal{O} \rangle =\frac{1}{1+e^{-2 \pi m}} \langle [X, \mathcal{O}] \rangle,  \qquad \qquad  \langle Y \mathcal{O} \rangle =\frac{1}{1+e^{2 \pi m}} \langle [Y, \mathcal{O}]\rangle 
\end{equation}
which fixes all correlators as a sum over Wick contractions. 

The special sphere partition function is naturally normalized to 
\begin{equation}
Z_{\mathrm{hyper}}(m) = \frac{1}{2 \cosh \pi m} \equiv \frac{1}{\ch m}. 
\end{equation}
With this normalization, we have the expected
\begin{equation}
\langle \mu \rangle % = \left(\frac12 \frac{1}{1+e^{-2 \pi m}} -  \frac12 \frac{1}{1+e^{2 \pi m}} \right) \frac{1}{2 \cosh \pi m} 
= \frac{\sinh \pi m}{4 \cosh^2 \pi m} = - \frac{1}{2 \pi} \partial_m \langle 1 \rangle. 
\end{equation}
This type of relation holds for all correlation functions of $U(1)$-invariant operators. 

The twisted trace necessarily agrees in appropriate chambers with the twisted trace over highest weight modules for the Weyl algebra 
generated by $X$ and $Y$. For example, we can expand for positive $m$ 
\be
\langle 1 \rangle= \sum_{n\geq 0} (-1)^n e^{- 2\pi (n+\frac12) m}
\ee
and 
\be
\langle X Y \rangle = \sum_{n\geq 0} (n+1) (-1)^n e^{- 2\pi (n+\frac12) m}, \qquad \qquad  \langle Y X \rangle =\sum_{n\geq 0} n (-1)^n e^{- 2\pi (n+\frac12) m}
\ee
which is the twisted trace on the highest weight module with basis $y^n$, with $X$ acting as $\partial_y$ and $Y$ as multiplication by $y$,
with fugacity $e^{2 \pi m}$ for $X$. 

We can also expand for negative $m$ as 
\be
\langle 1 \rangle= \sum_{n\geq 0} (-1)^n e^{2\pi (n+\frac12) m}
\ee
and 
\be
\langle X Y \rangle = - \sum_{n\geq 0} n (-1)^n e^{2\pi (n+\frac12) m}, \qquad \qquad  \langle Y X \rangle =- \sum_{n\geq 0} (n+1) (-1)^n e^{2\pi (n+\frac12) m}
\ee
which is the twisted trace on the highest weight module with basis $x^n$, with $X$ acting as multiplication by $x$ and $Y$ as $- \partial_x$,
with fugacity $e^{2 \pi m}$ for $X$. 

\section{Abelian examples}
\label{sec_abelian}
%%%%%%%%%%%%%%%%%%%%%%%%%%%%%%%%%%%%%%%%%%%
\subsection{SQED$_1$}
\label{sec_sqed1}
%%%%%%%%%%%%%%%%%%%%%%%%%%%%%%%%%%%%%%%%%%%
The $S^3$ partition function is
\be
Z_{\mathrm{SQED}_1} = \int_{-\infty}^\infty dx \frac{e^{2 \pi i x \zeta}}{2 \cosh \pi x} = \frac{1}{2 \cosh \pi \zeta}. 
\ee
This is obviously consistent with mirror symmetry to a single free hyper. 

This theory has trivial Higgs branch algebra. The $U(1)$ quantum Hamiltonian reduction of the Weyl algebra is trivial, 
as all $U(1)$ invariant operators are polynomials in the moment map. 
 
``Quantum''  Coulomb branch operators are generated by the scalar $z$ and 
Abelian monopoles $v_\pm$, with 
\begin{align}
v_\pm z &= (z \pm 1) v_{\pm}, \cr
v_+ v_- &= z+\frac12, \cr
v_- v_+ &= z-\frac12. 
\end{align} 
This is just the Weyl algebra. 

As $v_\pm$ have flavour charge $\pm 1$, the only non-zero twisted traces can be $\langle z^n \rangle$.
We have a relation 
\be
\langle p(z) (z-\frac12) \rangle = \langle p(z) v_- v_+ \rangle = -e^{- 2 \pi \zeta} \langle v_+ p(z) v_- \rangle = -e^{- 2 \pi \zeta} \langle p(z+1) (z+\frac12) \rangle
\ee
which allows one to compute them recursively. For example, 
\be
(1+e^{- 2 \pi \zeta}) \langle z \rangle = \frac12(1-e^{- 2 \pi \zeta}) \langle 1 \rangle
\ee
matches the mirror calculation for $\langle \mu \rangle$. 

The recursion relation is naturally satisfied by the integral formula 
\be
\langle p(z) \rangle= \int_{-\infty}^\infty dx \frac{e^{2 \pi i x \zeta}}{2 \cosh \pi x} p(-i x)
\ee
as
\be
\int_{-\infty}^\infty dx \frac{e^{2 \pi i x \zeta}}{2 \cosh \pi x} p(-i x) (-i x-\frac12) =- e^{- 2 \pi \zeta} \int_{-\infty}^\infty dx \frac{e^{2 \pi i x \zeta}}{2 \cosh \pi x} p(-i x+1) (-i x+\frac12) 
\ee
where the shift of $x$ by $i$ does not lead the contour to catch any poles: the dangerous $\cosh$ pole at $x=\frac{i}{2}$ is cancelled by the explicit factor of $(-i x-\frac12)$.

The integral can be computed by taking $\zeta$ to be positive or negative and deforming the contour to a sum over residues at
$-i x = n +\frac12$. This sum over residues also has an interpretation as a sum over equivariant vortex configurations in the only vacuum of the theory. 

More concretely, the sum over the residues at positive or negative $-i x = n +\frac12$ with an insertion of $p(z)$ gives sums of the form $\sum_n (-1)^n e^{-2 \pi (n+\frac12) \zeta} p(n+\frac12)$ 
which are obviously twisted traces of $p(z)$ over the highest or lowest weight modules for the quantum Coulomb branch algebra. 

%%%%%%%%%%%%%%%%%%%%%%%%%%%%%%%%%%%%%%%%%%%
\subsection{SQED$_2$}
\label{sec_sqed2}
%%%%%%%%%%%%%%%%%%%%%%%%%%%%%%%%%%%%%%%%%%%
The $S^3$ partition function is
\be
Z_{\mathrm{SQED}_2} = \int_{-\infty}^\infty dx \frac{e^{2 \pi i x \zeta}}{\ch \, x \,\ch \, (x+m)}. 
\ee

For positive $\zeta$, we can pick two towers of poles:
\begin{itemize}
\item At $ x = -m + (n + \frac12) i$ we get 
\be
Z^{(1)}_{\mathrm{SQED}_2} = -\sum_{n \geq 0} \frac{e^{- 2 \pi (n + \frac12) \zeta - 2 \pi i m}}{2 i \sinh \pi m} = i \frac{e^{-2 \pi i m \zeta}}{4 \sinh \pi m \sinh \pi \zeta}. 
\ee
\item At $ x = (n + \frac12) i$ we get 
\be
Z^{(2)}_{\mathrm{SQED}_2} = -i \frac{1}{4 \sinh \pi m \sinh \pi \zeta}. 
\ee
\end{itemize}

The combination 
\be
Z_{\mathrm{SQED}_2} = Z^{(1)}_{\mathrm{SQED}_2} + Z^{(2)}_{\mathrm{SQED}_2}
\ee
is better behaved than the two individual terms, as it is non-singular for $m\to0$ and for $\zeta \to 0$.

We identify the two terms to the two vacua of the theory, where either of the two hypermultiplets gets a vev to 
satisfy the moment map relations, and one has to set either $\vec \phi = 0$ or $\vec \phi + \vec m=0$. 
The flavour moment maps are then either $0$ or $-\vec \zeta$, leading to central charges $0$ or $-\vec m \vec \zeta$.

This matches the above exponential prefactors. 

\subsubsection{Higgs correlators}

The quantum Hamiltonian reduction enforces 
\begin{equation}
X_1 Y_1 + Y_1 X_1 +  X_2 Y_2 + Y_2 X_2= 2 i \zeta. 
\end{equation}
The Higgs branch algebra is generated by the remaining mesons:
\begin{equation}
F= X_1 Y_2, \qquad \qquad H=X_1 Y_1 - X_2 Y_2, \qquad \qquad E= X_2 Y_1
\end{equation}
with 
\begin{equation}
[F,H] = 2 F, \qquad \qquad [E,H] = - 2 E, \qquad \quad [E,F]=H
\end{equation}
and 
%\begin{equation}
%(H + i \zeta +\frac12)(-H + i \zeta -\frac12) = 4 X_1 Y_1 Y_2 X_2 = F E
%\end{equation}
%i.e. 
%\begin{equation}
%(H + i \zeta -\frac12)(-H + i \zeta +\frac12) = 4 X_1 Y_1 Y_2 X_2 = E F
%\end{equation}
\begin{align}
(H+i\zeta+1)(-H+i\zeta-1)&=4X_{1}Y_{1}Y_{2}X_{2}=4FE
\end{align}
i.e. 
\begin{align}
(H+i\zeta-1)(-H+i\zeta+1)&=4X_{2}Y_{2}Y_{1}X_{1}=4EF. 
\end{align}
It is the quotient of the universal enveloping algebra $U(\mathfrak{sl}_2)$ by a constraint fixing the Casimir to $-\frac14 (\zeta^2 + 1)$,
i.e. the spin to $-\frac12 \pm \frac{i}{2} \zeta$.

The mass $m$ is the fugacity for the flavour symmetry generated by 
\begin{equation}
\frac12 (X_2 Y_2 + Y_2 X_2) = \frac12(i \zeta -H). 
\end{equation}

The quantum Higgs branch algebra has two Verma modules, highest weight with spins $-\frac12 \pm \frac{i}{2} \zeta$,
which we associate to the two vacua. 
The two vacua of the theory should correspond to the two Verma modules for the 
quantum Higgs branch algebra. For positive mass, we can look at modules generated by a vector annihilated by 
$E$. That must have either $H=\pm i \zeta - 1$
%\frac12$ 
and thus fugacity power 
\begin{equation}
e^{- 2 \pi m q} = e^{- \pi m (i \zeta -H) } 
\end{equation}
which is either $e^{- \pi m}$ or $e^{- \pi m - 2\pi i m \zeta}$.
Every extra power of $F$ gives an extra factor of $e^{- 2 \pi m}$.

We can thus write 
\be
Z_{\mathrm{SQED}_2} =  i \frac{1}{2 \sinh \pi \zeta} \chi^H_1(m;\zeta)-i \frac{1}{2 \sinh \pi \zeta} \chi^H_2(m;\zeta)
\ee
in terms of twisted characters $\chi_{i}^{H}(m;\zeta)$ of the Higgs branch Verma modules, as expected. 

We expect Higgs correlators to admit a similar decomposition:
\be
\langle \mathcal{O}_H \rangle_{S^3} =  i \frac{1}{2 \sinh \pi \zeta} \langle \mathcal{O}_H \rangle_{1} -i \frac{1}{2 \sinh \pi \zeta} \langle \mathcal{O}_H \rangle_{2}. 
\ee
The coefficients here have the same $\zeta$-dependence, but this does not need to be true in general. It is due to the fact that the theory 
has a symmetry exchanging the two vacua. 

\subsubsection{Coulomb correlators}

The Coulomb branch algebra is now generated by 
\begin{align}
v_\pm z &= (z \pm 1) v_{\pm}, \cr
v_+ v_- &= (z+\frac12)(z-i m+\frac12), \cr
v_- v_+ &= (z-\frac12)(z-i m-\frac12). 
\end{align} 
Highest or lowest weight Verma modules can be built from a vector annihilated by $v_-$ or from a vector annihilated by $v_+$.

A vector annihilated by $v_-$ must have eigenvalues $z=-\frac12$ or $z = i m - \frac12$. The action of $v_+$ further lowers the 
$z$ eigenvalue, so the twisted trace involves a sum over terms weighted by $e^{2 \pi \zeta (n+\frac12)}$ or $e^{-2 \pi i m \zeta+ 2 \pi \zeta (n+\frac12)}$.

Similarly, a vector annihilated by $v_+$ must have eigenvalues $z=+\frac12$ or $z = i m + \frac12$. The action of $v_-$ further raises the 
$z$ eigenvalue, so the twisted trace involves a sum over terms weighted by $e^{-2 \pi \zeta (n+\frac12)}$ or $e^{-2 \pi i m \zeta- 2 \pi \zeta (n+\frac12)}$.

This is all consistent with the sum over residues, so we can write consistently something like 
\be
Z_{\mathrm{SQED}_2} =\frac{i}{2 \sinh \pi m} \chi^C_1(\zeta;m) - \frac{i}{2 \sinh \pi m} \chi^C_2(\zeta;m) 
\ee
in terms of twisted characters $\chi_{i}^{C}(\zeta;m)$ of the Coulomb branch Verma module 
and we also have 
\be
\langle p(z) \rangle =\frac{i}{2 \sinh \pi m} \langle p(z) \rangle_{V_1} - \frac{i}{2 \sinh \pi m} \langle p(z) \rangle_{V_2}. 
\ee

%%%%%%%%%%%%%%%%%%%%%%%%%%%%%%%%%%%%%%%%%%%
\subsection{SQED$_3$}
\label{sec_sqed3}
%%%%%%%%%%%%%%%%%%%%%%%%%%%%%%%%%%%%%%%%%%%
The $S^3$ partition function is
\begin{align}
Z_{\mathrm{SQED}_3} = 
\int_{-\infty}^{\infty} dx \frac{e^{2\pi ix\zeta}}{\ch x \, \ch (x+m_{1}) \, \ch (x+m_{2})}. 
\end{align}

For positive $\zeta$, one can choose three towers of poles:
\begin{itemize}
\item At $x=\left(n+\frac12\right)i$ we obtain
\begin{align}
\label{sqed3_1}
Z_{\mathrm{SQED}_3}^{(1)} = 
-\frac{1}{\sh m_{1} \sh m_{2} \ch \zeta}. 
\end{align}

\item At $x=-m_{1}+\left(n+\frac12\right)i$ we obtain
\begin{align}
\label{sqed3_2}
Z_{\mathrm{SQED}_3}^{(2)} = 
-\frac{e^{-2\pi i\zeta m_{1}}}{\sh m_{1} \sh(m_{1}-m_{2}) \ch\zeta}. 
\end{align}

\item At $x=-m_{2}+\left(n+\frac12\right)i$ we obtain
\begin{align}
\label{sqed3_3}
Z_{\mathrm{SQED}_3}^{(3)} = 
-\frac{e^{-2\pi i\zeta m_{2}}}{\sh m_{2}\sh(m_{2}-m_{1}) \ch\zeta}. 
\end{align}
\end{itemize}

The sum 
\be
\label{sqed3_sum}
Z_{\mathrm{SQED}_3} =Z_{\mathrm{SQED}_3}^{(1)}+Z_{\mathrm{SQED}_3}^{(2)}+Z_{\mathrm{SQED}_3}^{(3)}
\ee
is much better behaved than the individual terms. For example, it is finite as $m_{i}\rightarrow 0$ or $\zeta \to \pm \frac{i}{2}$. 

The three towers of poles correspond to the three vacua of the theory. 
The three exponential prefactors $1$, $e^{-2\pi i\zeta m_{1}}$, $e^{-2\pi i\zeta m_{2}}$ describe 
the central charges in the vacua. 

\subsubsection{Higgs correlators}
When we compute the correlation functions of Higgs branch local operators, 
the quantum Hamiltonian reduction requires that 
\begin{align}
\sum_{i=1}^{3}X_{i}Y_{i}+Y_{i}X_{i}&=2i\zeta.
\end{align}
The quantized Higgs branch algebra is identified with 
a quotient of the universal enveloping algebra of $\mathfrak{sl}_{3}$ 
whose Chevalley-Serre generators are
\begin{align}
\label{sqed3_AH1}
F_{1}&=X_{1}Y_{2},& H_{1}&=X_{1}Y_{1}-X_{2}Y_{2},& E_{1}&=X_{2}Y_{1},\nonumber\\
F_{2}&=X_{2}Y_{3},& H_{2}&=X_{2}Y_{2}-X_{3}Y_{3},& E_{2}&=X_{3}Y_{2},\nonumber\\
F_{3}&=[F_{2},F_{1}]=X_{1}Y_{3},& E_{3}&=[E_{1},E_{2}]=X_{3}Y_{1}.
\end{align}
They obey
\begin{align}
\label{sqed3_AH2}
[F_{i},H_{j}]&=2F_{i}\delta_{ij},& 
[E_{i},H_{j}]&=-2E_{i}\delta_{ij},& 
[E_{i},F_{j}]&=H_{i}\delta_{ij},\qquad i=1,2
\end{align}
and
\begin{align}
\label{sqed3_AH3}
(2H_{1}+H_{2}+i\zeta+\frac32)(-H_{1}+H_{2}+i\zeta-\frac32)&=9X_{1}Y_{1}Y_{2}X_{2}=9F_{1}E_{1},\nonumber\\
(-H_{1}+H_{2}+i\zeta+\frac32)(-H_{1}-2H_{2}+i\zeta-\frac32)&=9X_{2}Y_{2}Y_{3}X_{3}=9F_{2}E_{2},\nonumber\\
(2H_{1}+H_{2}+i\zeta+\frac32)(-H_{1}-2H_{2}+i\zeta-\frac32)&=9X_{1}Y_{1}Y_{3}X_{3}=9F_{3}E_{3}.
\end{align}
The masses $m_{1}$ and $m_{2}$ are associated with the flavour charges 
\begin{align}
q_{1}&=\frac12 (X_{2}Y_{2}+Y_{2}X_{2})=-\frac13(H_{1}-H_{2})+\frac{i\zeta}{3},
\nonumber\\
q_{2}&=\frac12 (X_{3}Y_{3}+Y_{3}X_{3})=-\frac13(H_{1}+2H_{2})+\frac{i\zeta}{3}. 
\end{align}

Three vacua of the theory admit the three Verma modules for the quantized Higgs branch algebra. 
For positive masses $m_{1}$ and $m_{2}$, the modules can be built from a vector annihilated by $E_{1}$, $E_{2}$ and $E_{3}$.  
It should have either $(q_{1},q_{2})$ $=$ $(\frac12,\frac12)$ or $(-\frac12, i\zeta+1)$ or $(i\zeta,\frac12)$ 
and the fugacity power 
\begin{align}
e^{-2\pi (m_{1}q_{1}+m_{2}q_{2})}
&=e^{\frac{2\pi m_{1}(H_{1}-2H_{2})}{3}-\frac{2\pi im_{1}\zeta}{3}+\frac{2\pi m_{2}(H_{1}+2H_{2})}{3}-\frac{2\pi im_{2}\zeta}{3}}
\end{align}
is either $e^{-\pi(m_{1}+m_{2})}$ or $e^{\pi(m_{1}-m_{2})-2\pi im_{2}\zeta}$ or $e^{-\pi m_{2}-2\pi im_{1}\zeta}$. 
By acting with $F_{1}$ and $F_{2}$, an extra factor of $e^{-2\pi m_{1}}$ and $e^{-2\pi m_{2}}$ appear. 

Thus we can write the $S^3$ partition function 
\begin{align}
Z_{\mathrm{SQED}_3}&=
-\frac{1}{\ch \zeta} \chi_{1}^{H}(m_{i};\zeta)
-\frac{1}{\ch \zeta} \chi_{2}^{H}(m_{i};\zeta)
-\frac{1}{\ch \zeta} \chi_{3}^{H}(m_{i};\zeta)
\end{align}
in terms of the twisted characters $\chi^{H}_{i}(m_{i};\zeta)$ of the Verma modules for the Higgs branch algebra. 

\subsubsection{Coulomb correlators}
We can address the correlation function of Coulomb branch operators 
from the Coulomb branch algebra
\begin{align}
\label{sqed3_Calg}
v_{\pm}z&=(z\pm1)v_{\pm},\nonumber\\
v_{+}v_{-}&=(z+\frac12)(z-im_{1}+\frac12)(z-im_{2}+\frac12),\nonumber\\
v_{-}v_{+}&=(z-\frac12)(z-im_{1}-\frac12)(z-im_{2}-\frac12).
\end{align}

Highest (resp. lowest) weight Verma modules can be constructed from a vector 
annihilated by $v_{-}$ (resp. $v_{+}$) with the eigenvalues $z=-\frac12$ or $im_{a}-\frac12$ (resp. $z=\frac12$ or $im_{a}+\frac12$). 
The action of $v_{+}$ (resp. $v_{-}$) decreases (resp. increases) the $z$ eigenvalues 
so that the twisted trace is a sum over terms weighted by 
$e^{2\pi \zeta(n+\frac12)}$ or $e^{-2\pi i m_{i}\zeta+2\pi \zeta (n+\frac12)}$ 
(resp. $e^{-2\pi \zeta(n+\frac12)}$ or $e^{-2\pi i m_{i}\zeta-2\pi \zeta (n+\frac12)}$ ). 

Hence we can write
\begin{align}
Z_{\mathrm{SQED}_3}&=
-\frac{1}{\sh m_{1}\sh m_{2}}\chi_{1}^{C}(\zeta;m_{i})
-\frac{1}{\sh m_{1}\sh (m_{1}-m_{2})}\chi_{2}^{C}(\zeta;m_{i})
\nonumber\\
&+\frac{1}{\sh m_{2}\sh (m_{1}-m_{2})}\chi_{3}^{C}(\zeta;m_{i})
\end{align}
where $\chi_{i}^{C}(\zeta;m_{i})$ are the twisted characters of the Verma module of quantum Coulomb branch algebra. 

%%%%%%%%%%%%%%%%%%%%%%%%%%%%%%%%%%%%%%%%%%%
\subsection{General Abelian theory}
\label{sec_gAbe}
%%%%%%%%%%%%%%%%%%%%%%%%%%%%%%%%%%%%%%%%%%%
Now consider the general Abelian gauge theory 
with gauge group $G=\prod_{i=1}^{r}U(1)_{i}$ 
and $N$ hypermultiplets $(X^{A},Y^{A})$ with $A=1,\cdots, N$ carrying charges $(Q^{i}_{A}, -Q^{i}_{A})$ under the $U(1)_{i}$. 
The theory has flavour symmetry $G_{H}$ $=$ $\prod_{a=1}^{N-r} U(1)_{a}$ 
that rotates the hypermultiplets $(X^{A},Y^{A})$ with charges $(q^{a}_{A},-q^{a}_{A})$. We assume all charges to be integral.

The topological symmetry of the theory is classically $G_{C}$ $=$ $U(1)^{r}$ that shifts the dual photons. 
We define a square $N\times N$ matrix 
\begin{align}
\label{charge_mtx}
\mathbf{Q}&=\left(
\begin{matrix}
Q&q\\
\end{matrix}
\right)
\end{align}
so that 
$\mathbf{Q}^{I}_{A}$ is $Q^{I}_{A}$ for $I=1,\cdots, r$ 
and $q^{I-r}_{A}$ for $I=r+1,\cdots, N$. 
We can introduce $(N-r)$ mass parameters $m^{a}_{A}$ 
for each $U(1)_{a}$ factor in the flavour symmetry group $G_{H}=U(1)^{N-r}$. 

For example, the matrix (\ref{charge_mtx}) for SQED$_{N_{f}}$ is
\begin{align}
\label{charge_sqed}
\mathbf{Q}&=\left(
\begin{matrix}
1&0&0&\cdots&0\\ 
1&1&0&\cdots&0\\
1&0&1&\cdots&0\\
1&0&0&\ddots&0\\
1&0&0&\cdots&1\\
\end{matrix}
\right).
\end{align}

In a massive trivial vacuum, a collection of $r$ hypermultiplets gain a vev. The gauge symmetry is completely Higgsed iff the restriction $\hat Q$ of the charge matrix to the hypers which gain a vev 
has determinant $1$. We assume the theory has massive trivial vacua only. 

The $S^3$ partition function takes the form:
\begin{align}
\label{generalAbe}
Z_{\textrm{Abelian}}&=
\prod_{i=1}^{r}\int_{-\infty}^{\infty} dx_{i} 
\frac{e^{2\pi i\sum_{i=1}^{r} x_{i}\zeta_{i}}}
{2^N \prod_{A=1}^{N} \cosh \pi (\sum_{i=1}^{r}Q_{A}^{i}x_{i}-\sum_{a=1}^{N-r}q_{A}^{a}m_{a})}. 
\end{align}
The factor $\cosh\pi(\sum_{i=1}^{r} Q^{i}_{A}x_{i}+\sum_{a=1}^{N-r}q_{A}^{a}m_{a})$ 
in the denominator corresponds to the $A$-th hypermultiplet. 

Mirror symmetry of Abelian theories is implemented in the special sphere partition function simply by a Fourier transform of the $\cosh$ factors, to 
\begin{align}
\label{generalAbe}
Z_{\textrm{Abelian}}&=
\prod_{i=1}^{r}\int_{-\infty}^{\infty} dx_{i} \prod_{A=1}^N \int_{-\infty}^{\infty} ds_{A} 
\frac{e^{2\pi i\sum_{i=1}^{r} x_{i}\zeta_{i} - 2 \pi i \sum_{A=1}^N s_A (\sum_{i=1}^{r}Q_{A}^{i}x_{i}-\sum_{a=1}^{N-r}q_{A}^{a}m_{a})}}
{2^N \prod_{A=1}^{N} \cosh \pi s_A}. 
\end{align}
and then doing the $x_i$ integrals to produce $\delta$ functions.

Once we give $\zeta_i$, say, a positive imaginary part and close the contour at infinity, the relevant residues indeed come in families labelled by the massive vacua. 
We can pick a vacuum and decompose the matrices $Q$ and $q$ as
\begin{align}
\label{Q_minor}
Q&=\left(
\begin{matrix}
\widehat{Q}\\
\widetilde{Q}\\
\end{matrix}
 \right),& 
 q&=\left(
\begin{matrix}
\widehat{q}\\
\widetilde{q}\\
\end{matrix}
 \right)
\end{align}
where $\widehat{Q}$ and $\widetilde{Q}$ are $r\times r$ and $(N-r)\times r$ matrices 
which encode the gauge charges for the hypers which get vevs and those for the others respectively. 
Similarly, $\widehat{q}$ are the $r\times (N-r)$ and $(N-r)\times (N-r)$ matrices 
which describe the flavour charges for the hypers with vevs and for the others respectively. 
The matrix $\widehat{Q}$ corresponds to a certain collection of hypermultiplets which get vevs. 

The residues sit at 
\begin{equation}
\left(\widehat{Q} x - \widehat{q} m \right)_j = (n_j + \frac12)i. 
\end{equation}
Because of the assumption $\det \widehat{Q}=1$, the inverse $\widehat{Q}^{-1}$ is a matrix of integers and thus the locations of the poles 
in the family differ by integer multiples of $i$. The residues will differ at most by a sign and a factor $\exp 2 \pi i \zeta  \cdot \widehat{Q}^{-1} n$.

Furthermore, $\det \widehat{Q}=1$ also insures the absence of a Jacobian factor in evaluating the residue. The residues within each family 
will thus resum to 
\begin{align}
\label{general_res}
Z_{\textrm{Abelian}}^{\widehat{Q}}
&=\frac{1}{2^{N}} \frac{
e^{2\pi i\widehat{\zeta}^{T}\widehat{Q}^{-1}\widehat{q}\mathbf{m}}
}
{
\prod_j \frac{
\left(
e^{\pi (\zeta \cdot \widehat{Q}^{-1})_j}
+(-1)^{\sum_a(\widetilde{Q}\widehat{Q}^{-1})^a_j}
e^{-\pi (\zeta \cdot \widehat{Q}^{-1})_j}
\right)}{2} \prod_a
\cosh\pi \left[
\left(
\widetilde{Q}\widehat{Q}^{-1}\widehat{q}
-\widetilde{q}
\right)\mathbf{m}
+\sum_j \frac{i}{2}( \widetilde{Q}\widehat{Q}^{-1})^j
\right]_a
}.
\end{align}

%%%%%%%%%%%%%%%%%%%%%%%%%%%%%%%%%%%%%%%%%%%
%%%%%%%%%%%%%%%%%%%%%%%%%%%%%%%%%%%%%%%%%%%
\section{SQCD}
\label{sec_sqcd}
%%%%%%%%%%%%%%%%%%%%%%%%%%%%%%%%%%%%%%%%%%%
%%%%%%%%%%%%%%%%%%%%%%%%%%%%%%%%%%%%%%%%%%%
Consider SQCD with gauge group $G=U(N_{c})$ and $N_{f}$ hypermultiplets $(X^{\alpha}_{i}, Y^{i}_{\alpha})$ 
with $i=1,\cdots, N_{c}$ and $\alpha=1,\cdots, N_{f}$ in the fundamental representation of the gauge group, i.e. 
$\mathcal{R}$ $\simeq$ $(T^{*}\mathbb{C}^{N_{c}})^{\oplus N_{f}}$ $\simeq$ $T^{*}\mathbb{C}^{N_{c}N_{f}}$. 
It has a topological symmetry $G_{C}=U(1)_{t}$ and a Higgs branch flavour symmetry $G_{H}=PSU(N_{f})$. 

The massive vacua are associated to a vev for $N_c$ out of $N_f$ hypers, Higgsing the gauge group completely. There are ${N_f \choose N_c}$ of them.

The $S^3$ partition function is given by
\begin{align}
\label{sqcd_s3}
Z_{(N_{c})-[N_{f}]}
&=
\frac{1}{N_{c}!}
\int \prod_{i=1}^{N_{c}}dx_{i}
\frac{\prod_{i<j}\sh^2 (x_{i}-x_{j}) e^{2\pi i\zeta\sum_{i=1}^{N_{c}} x_{i}}}
{\prod_{i=1}^{N_{c}} \prod_{\alpha=1}^{N_{f}} \ch (x_{i}-m_{\alpha})}. 
\end{align}

For good and ugly theories classified in \cite{Gaiotto:2008ak}, the matrix model (\ref{sqcd_s3}) is convergent for any real FI parameter $\zeta$. 
When $\zeta$ is positive, we can choose poles at
\begin{align}
\label{sqcd_pole}
x_{l}&=m_{i_{l}}+\left(n_{l}+\frac12 \right)i
\end{align}
where the integers $i_{l}$ with $l=1,\cdots, N_{c}$ should be distinct to avoid vanishing Vandermonde factors. We can complete them to a permutation of $N_{f}$ integers
\begin{align}
\label{sqcd_pole2}
(i_{1},i_{2},\cdots, i_{N_{c}},\cdots, i_{N_{f}})
&=(\sigma(1),\sigma(2),\cdots, \sigma(N_{c}), \cdots, \sigma(N_{f})). 
\end{align}
in some arbitrary way. 

From the poles at (\ref{sqcd_pole}) we obtain the contribution 
\begin{align}
\label{sqcd_s3sub}
Z_{(N_{c})-[N_{f}]}^{(\sigma)}
&=
\frac{
i^{-N_{c}(N_{f}+1)} e^{2\pi i\zeta \sum_{j=1}^{N_{c} }m_{\sigma(j)}}
}
{
\left(
e^{\pi \zeta}+(-1)^{N_{f}-1} e^{-\pi \zeta}
\right)^{N_{c}}
\prod_{j=1}^{N_{c}}\prod_{k=N_{c}+1}^{N_{f}}
\sh (m_{\sigma(j)}-m_{\sigma(k)})
}. 
\end{align}

As a result, the $S^3$ partition function is expressed as a sum over the vacua \cite{Kapustin:2010mh}
\begin{align}
\label{sqcd_sum}
Z_{(N_{c})-[N_{f}]}
&=
\sum_{\sigma\in S_{N_{f}}/(S_{N_{c}} \times S_{N_{f}-N_c})}
\frac{
i^{-N_{c}(N_{f}+1)} e^{2\pi i\zeta \sum_{j=1}^{N_{c} }m_{\sigma(j)}}
}
{
\left(
e^{\pi \zeta}+(-1)^{N_{f}-1} e^{-\pi \zeta}
\right)^{N_{c}}
\prod_{j=1}^{N_{c}}\prod_{k=N_{c}+1}^{N_{f}}
\sh (m_{\sigma(j)}-m_{\sigma(k)})
}. 
\end{align}

%%%%%%%%%%%%%%%%%%%%%%%%%%%%%%%%%%%%%%%%%%%
%%%%%%%%%%%%%%%%%%%%%%%%%%%%%%%%%%%%%%%%%%%
\section{A richer non-Abelian example}
\label{sec_TG}
%%%%%%%%%%%%%%%%%%%%%%%%%%%%%%%%%%%%%%%%%%%
%%%%%%%%%%%%%%%%%%%%%%%%%%%%%%%%%%%%%%%%%%%

%%%%%%%%%%%%%%%%%%%%%%%%%%%%%%%%%%%%%%%%%%%
\subsection{$T[SU(N)]$}
\label{sec_TSUN}
%%%%%%%%%%%%%%%%%%%%%%%%%%%%%%%%%%%%%%%%%%%
The $T[SU(N)]$ originally introduced in \cite{Gaiotto:2008ak} is a linear quiver gauge theory 
with a gauge group $G=\prod_{\alpha=1}^{N-1} U(\alpha)$ 
and hypermultiplets transforming in bifundamental representations for all adjacent nodes. It has $N!$ vacua labelled by 
all possible ways to pair up masses and FI parameters.

The $S^3$ partition function takes the form 
\begin{align}
\label{tsuN_s3}
&Z_{T[SU(N)]}=\int dx^{(1)}\times \frac12 \int dx_{1}^{(2)}dx_{2}^{(2)}
\times \cdots \times \frac{1}{(N-1)!}\int \prod_{i=1}^{N-1}dx_{i}^{(N-1)}
\nonumber\\
&\times 
\frac{
\prod_{k=2}^{N-1}\prod_{i<j}^{k}\sh^{2}(x_{i}^{(k)}-x_{j}^{(k)})
e^{2\pi i(\zeta_{1}x^{(1)}+\zeta_{2}(x_{1}^{(1)}+x_{2}^{(2)})+\cdots+\zeta_{N-1}(x_{1}^{(N-1)}+\cdots+x_{N-1}^{(N-1)}) )} }
{\prod_{j=1}^{2}\ch (x_{j}^{(2)}-x^{(1)}) \cdots \prod_{i=1}^{k}\prod_{j=1}^{k-1} \ch(x_{i}^{(k)}-x_{j}^{(k-1)}) \cdots \prod_{i=1}^{N}\prod_{j=1}^{N-1} \ch(m_{i}-x_{j}^{(N-1)})}. 
\end{align}

For $\zeta_{i}>0$ we can pick up the poles at 
\begin{align}
\label{tsuN_pole}
x_{l}^{(k)}&=
m_{i_{l}}+\sum_{i=k}^{N-1}(n_{l}^{(i)}+\frac12)i, \qquad l=1,\cdots, k
\end{align}
where the integers $i_{l}$ with $l=1,\cdots, N-1$ are obtained by permuting $(N-1)$ integers:
\begin{align}
\label{tsuN_pole2}
(i_{1}, i_{2}, i_{3},\cdots, i_{N-1})&=(\sigma(1), \sigma(2), \sigma(3),\cdots, \sigma(N-1)). 
\end{align}
From the sequence of poles (\ref{tsuN_pole}) and their permutations 
the combinatorial factors in front of the integral (\ref{tsuN_s3}) are cancelled 
so that we get the contribution 
\begin{align}
\label{tsuN_s3sub}
Z_{T[SU(N)]}^{(\sigma)}
&=
\frac{(-1)^{\epsilon(\sigma)} e^{2\pi i\sum_{l=1}^{N-1} m_{i_{l}} (\zeta_{N-1}+\cdots+\zeta_{l})}}
{i^{\frac{N(N-1)}{2}}\prod_{i<j}\sh (m_{i}-m_{j}) \prod_{k=1}^{N-1}\prod_{l=1}^{k}\sh (\zeta_{k}+\zeta_{k-1}+\cdots+\zeta_{l})}. 
\end{align}

Since the sequence of poles (\ref{tsuN_pole}) is labelled by $\sigma \in S_{N}$, 
the $S^3$ partition function (\ref{tsuN_s3}) is expressed as a sum over the permutation group elements
\begin{align}
\label{tsuN_sum}
Z_{T[SU(N)]}&=
\sum_{\sigma\in S_{N}}
\frac{(-1)^{\epsilon(\sigma)} e^{2\pi i\sum_{l=1}^{N-1} m_{\sigma(l)} (\zeta_{N-1}+\cdots+\zeta_{l})}}
{i^{\frac{N(N-1)}{2}}\prod_{i<j}\sh (m_{i}-m_{j}) \prod_{k=1}^{N-1}\prod_{l=1}^{k}\sh (\zeta_{k}+\zeta_{k-1}+\cdots+\zeta_{l})}. 
\end{align}
By introducing variables 
\begin{align}
\zeta_{i}&=\widetilde{\zeta}_{i}-\widetilde{\zeta}_{i+1},& \sum_{i}^{N}\widetilde{\zeta}_{i}&=0,& i&=1,2,\cdots, N-1
\end{align}
the $S^3$ partition function (\ref{tsuN_sum}) takes the form of a sum over vacua
 \cite{Benvenuti:2011ga, Gulotta:2011si, Nishioka:2011dq} 
\begin{align}
\label{tsuN_sum2}
Z_{T[SU(N)]}&=
\sum_{\sigma\in S_{N-1}} 
\frac{(-1)^{\epsilon(\sigma)} e^{2\pi i\sum_{i=1}^{N}m_{\sigma(i)} (\widetilde{\zeta}_{i}-\widetilde{\zeta}_{N})}}
{i^{\frac{N(N-1)}{2}} \prod_{i<j} \sh (m_{i}-m_{j}) \prod_{i<j} \sh (\widetilde{\zeta}_{i}-\widetilde{\zeta}_{j})}. 
\end{align}

The quantum Higgs and Coulomb branch algebras are both quotients of the universal enveloping algebra of $\mathfrak{su}(N)$ 
by an ideal generated by the Casimirs. Correspondingly, they have $N!$ Verma modules, generated freely by the raising operators in the Lie algebra. 
We recognize this structure in the denominator factors above.  
%%%%%%%%%%%%%%%%%%%%%%%%%%%%%%%%%%%%%%%%%%%
\subsection{$T[G]$}
\label{sec_TG}
%%%%%%%%%%%%%%%%%%%%%%%%%%%%%%%%%%%%%%%%%%%
It has been proposed in \cite{Chang:2019dzt} that the $S^3$ partition function of $T[G]$ takes the form
\begin{align}
\label{TG_sum}
Z_{T[G]}&=\sum_{w\in \mathcal{W}(G)}
\frac{(-1)^{l(w)} e^{2\pi i (w(m)\cdot \zeta)}}
{i^{\frac{d}{2}} \prod_{\alpha\in \Delta_{+}} \sh(\alpha\cdot m) 
\prod_{\alpha\in \Delta_{+}} \sh(\alpha\cdot \zeta)}
\end{align}
where $l(w)$ is the length of an element $w$ in a Weyl group $\mathcal{W}(G)$ of $G$ 
and $d$ is the complex dimension of the Coulomb branch or equivalently that of the Higgs branch,
at least when the gauge algebra $\mathfrak{g}$ is self-Langlands dual. 

This generalizes the formula (\ref{tsuN_sum2}) of the $S^3$ partition function for $T[SU(N)]$, 
and again takes the form of a sum over products of character of Verma modules for the 
quantum Higgs and Coulomb branch algebra, which are both quotients of the universal enveloping algebra of $\mathfrak{g}$ 
by an ideal generated by the Casimirs.

%%%%%%%%%%%%%%%%%%%%%%%%%%%%%%%%%%%%%%%%%%%
%%%%%%%%%%%%%%%%%%%%%%%%%%%%%%%%%%%%%%%%%%%
\section{ADHM with one flavour}
\label{sec_ADHM}
%%%%%%%%%%%%%%%%%%%%%%%%%%%%%%%%%%%%%%%%%%%
%%%%%%%%%%%%%%%%%%%%%%%%%%%%%%%%%%%%%%%%%%%
Consider $N$ D2-branes which sit on top of a D6-brane in Type IIA string theory. 
The world-volume theory of the D2-branes is a 3d $\mathcal{N}=4$ $U(N)$ gauge theory 
described by the ADHM quiver with one adjoint hypermultiplet $(X,Y)$ and one fundamental hypermultiplet $(I,J)$. 
This is a self-mirror theory.

The vacua of the ADHM quiver are labelled by Young diagrams with $N$ boxes. The diagrams control the pattern of embedding of the $U(1)$ 
flavour symmetry into the Cartan of the gauge symmetry. The embedding is given by $N$ integers and the sequence of diagonal lengths $\ell_p$ of the Young diagram 
tells us that the embedding includes $\ell_p$ times the integer $p$. 

When looking for the corresponding collection of residues of the integral, that means that $\ell_p$ of the $x_*$ will take values which differ 
from $p m$ by some (half)integral multiple of $i$. The $\ell_0$ eigenvalues can give poles in denominator factors from the fundamental hyper.
Differences of $\ell_{p+1}$ and $\ell_p$ eigenvalues can give poles in denominator factors from the adjoint hyper. The Vandermonde 
factors can cancel some of the denominator factors and thus may require one to combine multiple denominator factors to get an overall simple pole. 

All of these constraints match the form of the Coulomb branch Verma modules, described i.e. in \cite{Gaiotto:2019wcc}, so that we have the expected one-to-one correspondence 
between residues and basis elements in the Verma module with weights $- i x_*$.

We will show now some explicit examples. 
%%%%%%%%%%%%%%%%%%%%%%%%%%%%%%%%%%%%%%%%%%%
\subsection{N=1}
\label{sec_ADHM1}
%%%%%%%%%%%%%%%%%%%%%%%%%%%%%%%%%%%%%%%%%%%
Up to a decoupled hyper, this is the same as the SQED with one flavour. The $S^3$ partition function is 
\begin{align}
\label{ADHM1}
Z_{\mathrm{ADHM}^1} = \int_{-\infty}^\infty dx \frac{e^{2 \pi i x \zeta}}{4 \cosh \pi m \cosh \pi x} = \frac{1}{\ch \,m \ch \zeta}
\end{align}
obviously self-mirror. 

%%%%%%%%%%%%%%%%%%%%%%%%%%%%%%%%%%%%%%%%%%%
\subsection{N=2}
\label{sec_ADHM2}
%%%%%%%%%%%%%%%%%%%%%%%%%%%%%%%%%%%%%%%%%%%
For $N=2$ the $S^3$ partition function takes the form
\begin{align}
\label{ADHM2}
Z_{\mathrm{ADHM}^2} =
\frac12 \int dx_{1}dx_{2}
\frac{\sh^2(x_{1}-x_{2}) e^{2\pi i\zeta (x_{1}+x_{2})}}
{\ch^{2}m \ \ch(x_{1}-x_{2}+m) \ch(x_{2}-x_{1}+m) \ch x_{1} \ch x_{2}}. 
%\frac12  \int_{-\infty}^\infty dx_1 dx_2
%\frac{1}{\ch }
%\frac{e^{2 \pi i x_1 \zeta}}{\ch \, x_1}\frac{e^{2 \pi i x_2 \zeta}}{\ch \, x_2}\frac{\sh^2 \, (x_1-x_2)}{ \ch\, (x_1-x_2+m)\ch\, (x_2-x_1+m)} 
\end{align}

We cannot get a pole from both $\ch \, x_1$ and $\ch \, x_2$, as that would cause a zero from the Vandermonde. We can instead have a pole from $\ch \, x_1$ 
and one from $\ch\, (x_2-x_1+m)$, or a pole from $\ch \, x_1$ and one from $\ch\, (x_2-x_1-m)$. The alternative $x_1 \leftrightarrow x_2$ is equivalent, and cancels the 
$\frac12$ combinatorial factor in front of the integral. 

The two contributions thus require $x_1 = i (n_1 + \frac12)$, $x_2 =- m +i (n_1 + \frac12)+i (n_2 + \frac12)$ (say with both summands positive) 
or $x_1 = i (n_1 + \frac12)$, $x_2 = m +i (n_1 + \frac12)+i (n_2 + \frac12)$. 
These sequences of poles correspond to the two Young diagrams $\tiny \yng(2)$ and $\tiny \yng(1,1)$ respectively. 

For the first sequence, the remaining integrand factors give $e^{2 \pi \zeta (- 2 (n_1 + \frac12) - (n_2 + \frac12)-i m)}$, an appropriate sign and a prefactor
$\frac{1}{\ch m\ch 2 m}$. Then the summation gives 
\begin{align}
\label{ADHM2_1}
Z_{\mathrm{ADHM}^2}^{\tiny \yng(2)}
=i\frac{e^{-2\pi i\zeta m}}{\sh 2m \cdot \ch m \cdot \sh 2\zeta \cdot \ch \zeta}.
\end{align}
The second series gives 
\begin{align}
\label{ADHM2_2}
Z_{\mathrm{ADHM}^2}^{\tiny \yng(1,1)}
=-i\frac{e^{2\pi i\zeta m}}{\sh 2m \cdot \ch m \cdot \sh 2\zeta \cdot \ch \zeta}. 
\end{align}
The sum 
\begin{align}
\label{ADHM2_sum}
Z_{\mathrm{ADHM}^2}&=
Z_{\mathrm{ADHM}^2}^{\tiny \yng(2)}+Z_{\mathrm{ADHM}^2}^{\tiny \yng(1,1)}
\nonumber\\
&=\frac{1}{32} \frac{\sin(2\pi m\zeta)}
{\sinh(\pi m)\cdot \cosh^{2}(\pi m) \sinh(\pi \zeta)\cosh^{2}(\pi \zeta)}
\end{align}
behaves nicely for $m\rightarrow 0$ and for $\zeta\rightarrow 0$ 
so that we have 
\begin{align}
\label{ADHM2_mt0}
Z_{\mathrm{ADHM}^2}&\xrightarrow{m,\zeta \rightarrow 0}\frac{1}{16 \pi}.
\end{align}

The $\zeta$-dependence automatically matches the Coulomb Verma module characters, and by mirror symmetry so does the $m$ dependence. 

%%%%%%%%%%%%%%%%%%%%%%%%%%%%%%%%%%%%%%%%%%%
\subsection{N=3}
\label{sec_ADHM3}
%%%%%%%%%%%%%%%%%%%%%%%%%%%%%%%%%%%%%%%%%%%
The $S^3$ partition function reads 
\begin{align}
\label{ADHM3}
Z_{\mathrm{ADHM}^3} = 
\frac{1}{3!}\int \prod_{i=1}^{3}dx_{i} 
\frac{\prod_{i<j} \sh^{2}(x_{i}-x_{j}) e^{2\pi i\zeta \sum_{i=1}^{3}x_{i}}}
{\prod_{i,j=1}^{3} \ch(x_{i}-x_{j}-m)\prod_{i=1}^{3}\ch x_{i}}.
\end{align}

There are three sequences of poles which contribute to the partition function, 
which correspond to the Young diagrams $\tiny \yng(3)$, $\tiny \yng(2,1)$ and $\tiny \yng(1,1,1)$. 
Under the permutation of $x_{i}$, 
the same contributions cancel the $\frac{1}{3!}$ combinatorial factor in the front of the integral. 
\begin{itemize}
\item 
At 
$x_{1}=i\left(n_{1}+\frac12\right)$, 
$x_{2}=-m+i\left(n_{1}+\frac12\right)+i\left(n_{2}+\frac12\right)$, 
$x_{3}=-2m+i\left(n_{1}+\frac12\right)+i\left(n_{2}+\frac12\right)+i\left(n_{3}+\frac12\right)$, we get 
\begin{align}
\label{ADHM3_1}
Z_{\mathrm{ADHM}^3}^{\tiny \yng(3)} = 
-i \frac{e^{-6\pi i\zeta m}}{\ch 3m \cdot \sh 2m \cdot \ch m
\cdot 
 \ch 3\zeta \cdot \sh 2\zeta \cdot \ch \zeta}.
\end{align}

\item 
At 
$x_{1}=i\left(n_{1}+\frac12\right)$, 
$x_{2}=-m+i\left(n_{1}+\frac12\right)+i\left(n_{2}+\frac12\right)$, 
$x_{3}=m+i\left(n_{1}+\frac12\right)+i\left(n_{3}+\frac12\right)$, we have
\begin{align}
\label{ADHM3_2}
Z_{\mathrm{ADHM}^3}^{\tiny \yng(2,1)} = 
\frac{1}{\ch 3m \cdot \ch^2 m \cdot \ch 3\zeta \cdot \ch^2 \zeta}. 
\end{align}

\item 
At 
$x_{1}=i\left(n_{1}+\frac12\right)$, 
$x_{2}=m+i\left(n_{1}+\frac12\right)+i\left(n_{2}+\frac12\right)$, 
$x_{3}=2m+i\left(n_{1}+\frac12\right)+i\left(n_{2}+\frac12\right)+i\left(n_{3}+\frac12\right)$, we obtain
\begin{align}
\label{ADHM3_3}
Z_{\mathrm{ADHM}^3}^{\tiny \yng(1,1,1)} = 
i \frac{e^{6\pi i\zeta m}}{\ch 3m \cdot \sh 2m \cdot \ch m \cdot \ch 3\zeta \cdot \sh 2\zeta \cdot \ch\zeta}. 
\end{align}
\end{itemize}

The combination 
\begin{align}
\label{ADHM3_sum}
Z_{\mathrm{ADHM}^3}&=
Z_{\mathrm{ADHM}^3}^{\tiny \yng(3)}+Z_{\mathrm{ADHM}^3}^{\tiny \yng(2,1)}+Z_{\mathrm{ADHM}^3}^{\tiny \yng(1,1,1)} 
\nonumber\\
&=
\frac{1}{64}
\frac{
\frac{1}{\cosh(\pi m) \cosh(\pi \zeta)}-\frac{2\sin(6\pi m\zeta)}{\sinh(2\pi m) \sinh(2\pi \zeta)}
}
{\cosh(\pi m)\cosh(3\pi m) \cosh(\pi \zeta)\cosh(3\pi \zeta)}
\end{align}
has a good behavior for $m\rightarrow 0$ and for $\zeta\rightarrow 0$ 
so that 
\begin{align}
\label{ADHM3_mt0}
Z_{\mathrm{ADHM}^3}&\xrightarrow{m,\zeta \rightarrow 0} \frac{\pi-3}{64 \pi}.
\end{align}

This is the first example where we have three vacua which are {\it not} related by some symmetry. It seems apparent that vacua are labelled by the same 
Young diagram on the Coulomb and Higgs sides. 

%%%%%%%%%%%%%%%%%%%%%%%%%%%%%%%%%%%%%%%%%%%
\subsection{N=4}
\label{sec_ADHM4}
%%%%%%%%%%%%%%%%%%%%%%%%%%%%%%%%%%%%%%%%%%%
Let us continue computation of the $S^3$ partition function for $N=4$. 
It is given by the integral
\begin{align}
\label{ADHM4}
Z_{\mathrm{ADHM}^4} = 
\frac{1}{4!}\int \prod_{i=1}^{4}dx_{i} 
\frac{\prod_{i<j} \sh^{2}(x_{i}-x_{j}) e^{2\pi i\zeta \sum_{i=1}^{4}x_{i}}}
{\prod_{i,j=1}^{4} \ch(x_{i}-x_{j}-m)\prod_{i=1}^{4}\ch x_{i}}.
\end{align}

There are five sequences of poles to be chosen. 
They are labelled by the Young diagrams 
$\tiny \yng(4)$, $\tiny \yng(3,1)$, $\tiny \yng(2,2)$, $\tiny \yng(2,1,1)$ and $\tiny \yng(1,1,1,1)$. 
For each of sequences 
permuting $x_{i}$, 
we get the same contributions which cancel the $\frac{1}{4!}$ combinatorial factor in the front of the integral. 
\begin{itemize}
\item At 
$x_{1}=i\left(n_{1}+\frac12\right)$, 
$x_{2}=-m+i\left(n_{1}+\frac12\right)+i\left(n_{2}+\frac12\right)$, 
$x_{3}=-2m+i\left(n_{1}+\frac12\right)+i\left(n_{2}+\frac12\right)+i\left(n_{3}+\frac12\right)$, 
$x_{4}=-3m+i\left(n_{1}+\frac12\right)+i\left(n_{2}+\frac12\right)+i\left(n_{3}+\frac12\right)+i\left(n_{4}+\frac12\right)$, 
we obtain
\begin{align}
\label{ADHM4_1}
Z_{\mathrm{ADHM}^4}^{\tiny \yng(4)} = 
-\frac{e^{-12\pi i\zeta m}}
{\sh 4m \cdot \ch 3m \cdot \sh 2m \cdot \ch m 
\cdot \sh 4\zeta \cdot \ch 3\zeta \cdot \sh 2\zeta \cdot \ch\zeta}. 
\end{align}

\item At 
$x_{1}=i\left(n_{1}+\frac12\right)$, 
$x_{2}=-m+i\left(n_{1}+\frac12\right)+i\left(n_{2}+\frac12\right)$, 
$x_{3}=-2m+i\left(n_{1}+\frac12\right)+i\left(n_{2}+\frac12\right)+i\left(n_{3}+\frac12\right)$, 
$x_{4}=m+i\left(n_{1}+\frac12\right)+i\left(n_{4}+\frac12\right)$,
we get 
\begin{align}
\label{ADHM4_2}
Z_{\mathrm{ADHM}^4}^{\tiny \yng(3,1)} = 
-\frac{e^{-4\pi i\zeta m}}
{\sh 4m \cdot \sh 2m \cdot \ch^2 m \cdot 
\sh 4\zeta \cdot \sh 2\zeta \cdot \ch^2 \zeta}. 
\end{align}

\item At 
$x_{1}=i\left(n_{1}+\frac12\right)$, 
$x_{2}=-m+i\left(n_{1}+\frac12\right)+i\left(n_{2}+\frac12 \right)$, 
$x_{3}=m+i\left(n_{1}+\frac12\right)+i\left(n_{3}+\frac12\right)$, 
$x_{4}=i\left(n_{2}+\frac12\right)+i\left(n_{3}+\frac12\right)+i\left(n_{4}+\frac12\right)$, 
we have 
\begin{align}
\label{ADHM4_3}
Z_{\mathrm{ADHM}^4}^{\tiny \yng(2,2)} = 
\frac{1}
{\ch 3m \cdot \sh^2 2m \cdot \ch m \cdot 
\ch 3\zeta \cdot \sh^2 2\zeta \cdot \ch \zeta}. 
\end{align}

\item At 
$x_{1}=i\left(n_{1}+\frac12\right)$, 
$x_{2}=-m+i\left(n_{1}+\frac12\right)+i\left(n_{2}+\frac12\right)$, 
$x_{3}=m+i\left(n_{1}+\frac12\right)+i\left(n_{3}+\frac12\right)$, 
$x_{4}=2m+i\left(n_{1}+\frac12\right)+i\left(n_{3}+\frac12\right)+i\left(n_{4}+\frac12\right)$, 
we find
\begin{align}
\label{ADHM4_4}
Z_{\mathrm{ADHM}^4}^{\tiny \yng(2,1,1)} = 
-\frac{e^{4\pi i\zeta m}}
{\sh 4m \cdot \sh 2m \cdot \ch^2 m \cdot 
\sh 4\zeta \cdot \sh 2\zeta \cdot \ch^2 \zeta}. 
\end{align}

\item At 
$x_{1}=i\left(n_{1}+\frac12\right)$, 
$x_{2}=m+i\left(n_{1}+\frac12\right)+i\left(n_{2}+\frac12\right)$, 
$x_{3}=2m+i\left(n_{1}+\frac12\right)+i\left(n_{2}+\frac12\right)+i\left(n_{3}+\frac12\right)$, 
$x_{4}=3m+i\left(n_{1}+\frac12\right)+i\left(n_{2}+\frac12\right)+i\left(n_{3}+\frac12\right)+i\left(n_{4}+\frac12\right)$, 
we obtain
\begin{align}
\label{ADHM4_5}
Z_{\mathrm{ADHM}^4}^{\tiny \yng(1,1,1,1)} = 
-\frac{e^{12\pi i\zeta m}}
{\sh 4m \cdot \ch 3m \cdot \sh 2m \cdot \ch m \cdot 
\sh 4\zeta \cdot \ch 3\zeta \cdot \sh 2\zeta \cdot \ch \zeta}.
\end{align}

\end{itemize}

The $S^3$ partition function is expressed as a sum 
\begin{align}
\label{ADHM4_sum}
Z_{\mathrm{ADHM}^4}&=
Z_{\mathrm{ADHM}^4}^{\tiny \yng(4)}+Z_{\mathrm{ADHM}^4}^{\tiny \yng(3,1)}+Z_{\mathrm{ADHM}^4}^{\tiny \yng(2,2)} 
+Z_{\mathrm{ADHM}^4}^{\tiny \yng(2,1,1)}+Z_{\mathrm{ADHM}^4}^{\tiny \yng(1,1,1,1)}
\nonumber\\
&=
\frac{1}{256}
\frac{1}{\sinh(2\pi m) \sinh(2\pi \zeta) \cosh(\pi m)\cosh(\pi \zeta)}
\nonumber\\
&\times 
\Biggl[
\frac{1}{\sinh(2\pi m) \sinh(2\pi \zeta) \cosh(3\pi m) \cosh(3\pi\zeta)}
\nonumber\\
&-\frac{2}{\sinh(4\pi m) \sinh(4\pi \zeta)} 
\left(
\frac{\cos(4\pi m\zeta)}{\cosh(\pi m) \cosh(\pi \zeta)}
+\frac{\cos(12\pi m\zeta)}{\cosh(3\pi m) \cosh(3\pi \zeta)}
\right)
\Biggr]
\end{align}
which is well-behaved for $m\rightarrow 0$ and for $\zeta\rightarrow 0$ 
so that 
\begin{align}
\label{ADHM4_mt0}
Z_{\mathrm{ADHM}^4}
&\xrightarrow{m,\zeta \rightarrow 0} 
%-\frac14+\frac{5}{2\pi^2}
-\frac{1}{1024}+\frac{5}{528\pi^2}. 
\end{align}

Again, it seems apparent that vacua are labelled by the same 
Young diagram on the Coulomb and Higgs sides. 

%%%%%%%%%%%%%%%%%%%%%%%%%%%%%%%%%%%%%%%%%%%
\subsection{General $N$}
\label{sec_ADHMN}
%%%%%%%%%%%%%%%%%%%%%%%%%%%%%%%%%%%%%%%%%%%
The $S^3$ partition function reads
\begin{align}
\label{ADHMN}
Z_{\mathrm{ADHM}^N} = 
\frac{1}{N!}\int \prod_{i=1}^{N}dx_{i} 
\frac{\prod_{i<j} \sh^{2}(x_{i}-x_{j}) e^{2\pi i\zeta \sum_{i=1}^{N}x_{i}}}
{\prod_{i,j=1}^{N} \ch (x_{i}-x_{j}-m)\prod_{i=1}^{N}\ch x_{i}}.
\end{align}
The poles of integrand can be specified by the Young diagram 
$\lambda$ $=$ $(\lambda_{1},\cdots,\lambda_{\lambda_{1}'})$ with $N$ boxes. 
For the poles encoded by the Young diagram we can relabel the gauge fugacities $x_{i}$ by
\begin{align}
\label{ADHMN_pole}
x_{k,a}&=
\begin{cases}
-km+\sum_{\alpha=1}^{a-1}\left(n_{\alpha,|k|+a}+\frac12\right)i+\sum_{\beta=1}^{|k|+a}\left(n_{a,\beta}+\frac12\right)i&\textrm{for $k\ge0$}\cr
-km+\sum_{\alpha=1}^{|k|+a-1}\left(n_{\alpha,a}+\frac12\right)i+\sum_{\beta=1}^{a}\left(n_{|k|+a,\beta}+\frac12\right)i&\textrm{for $k<0$}\cr
\end{cases}
\nonumber\\
&=
\begin{cases}
-km+\sum_{\alpha=1}^{a-1}n_{\alpha,|k|+a} i+\sum_{\beta=1}^{|k|+a}n_{a,\beta} i +\left(|k|+2a-1\right)\frac{i}{2}&\textrm{for $k\ge0$}\cr
-km+\sum_{\alpha=1}^{|k|+a-1}n_{\alpha,a} i+\sum_{\beta=1}^{a}n_{|k|+a,\beta} i +\left(|k|+2a-1\right)\frac{i}{2}&\textrm{for $k<0$}\cr
\end{cases}.
\end{align}
Here the integers $k$ $=$ $-(\lambda_{1}'-1), \cdots, -1,0,1,\cdots, (\lambda_{1}-1)$ index the diagonals of Young diagram 
and the positive integers $a$ $=$ $1, \cdots, n_{k}$ label the boxes in the $k$-th diagonal from the top left one 
with $n_{k}$ being the number of $k$-th diagonals of the Young diagram. 
The integer $n_{\alpha,\beta}=0,1,\cdots$ corresponds to the tower of poles associated to the box at $\alpha$-th row and $\beta$-th column. 

We find that the $S^3$ partition function (\ref{ADHMN}) is given by a sum over the Young diagrams $\lambda$:
\begin{align}
\label{ADHMN_sum}
Z_{\mathrm{ADHM}^N} = 
\sum_{\lambda}
\frac{C_{1}}{C_{2}C_{3}} 
\frac{e^{-2\pi i\zeta m \sum_{j}j n_{j}}}
{
\prod_{k=1}^{\infty}
\left( e^{k\pi m}-(-1)^{k}e^{-k\pi m} \right)^{d_{k}} \cdot \left( e^{k\pi \zeta}-(-1)^{k}e^{-k\pi \zeta} \right)^{d_{k}}} 
\end{align}
where 
\begin{align}
\label{coeff1}
C_{1}&=
\begin{cases}
-1&\textrm{for $N\equiv 2\mod 4$} \cr
1&\textrm{otherwise} \cr
\end{cases},
\\
\label{coeff2}
C_{2}&=\prod_{k=-(\lambda_{1}'-1)}^{\lambda_{1}-1}\prod_{\begin{smallmatrix}l=-(\lambda_{1}'-1)\\l\neq k-1\end{smallmatrix}}^{\lambda_{1}-1}
\prod_{a=1}^{n_{k}}\prod_{b=1}^{n_{l}}\left(\frac{k-l-1}{|k-l-1|} e^{-\frac{\pi i}{2}} \right)^{(|k|-|l|+2(a-b))},
\\
\label{coeff3}
C_{3}&=\prod_{\begin{smallmatrix}k=-(\lambda_{1}'-1)\\k\neq 0\\ \end{smallmatrix}}^{\lambda_{1}-1}
\prod_{a=1}^{n_{k}}\left(\frac{k}{|k|}e^{-\frac{\pi i}{2}}\right)^{|k|+2a-1},
\\
\label{power1}
d_{k}&=n_{k}+n_{-k}-\sum_{l}(n_{l}-n_{l-1})(n_{l+k}+n_{l-k}). 
\end{align}

Alternatively, the $S^3$ partition function (\ref{ADHMN_sum}) can be written as 
\begin{align}
\label{ADHMN_sum2}
Z_{\mathrm{ADHM}^N} &= 
\sum_{\lambda}
\frac{C_{1}}{C_{2}C_{3}} 
e^{\pi \zeta m \sum_{i}\lambda_{i}(-\lambda_{i}+2i-1)}
%e^{-2\pi i\zeta m \sum_{j}jn_{j}}
\prod_{b\in \lambda}
\frac{
e^{\pi (m+\zeta) h_{\lambda}(b)}}
{
\left(1-(-e^{2\pi m})^{h_{\lambda}(b)} \right)
\left(1-(-e^{2\pi \zeta})^{h_{\lambda}(b)} \right)
}
\end{align}
where we have rewritten the numerator by using the relation
\begin{align}
\label{diagonal_knk}
\sum_{k}k n_{k}&=\frac12 \sum_{i}\lambda_{i}(\lambda_{i}-2i+1). 
\end{align}
Here $h_{\lambda}(b)$ is the hook-length of a box $b$ at the $i$-th row and $j$-th column in a Young diagram $\lambda$ 
which is defined as the number of boxes below and to the right of $b$ including $b$ itself: 
\begin{align}
\label{hook}
h_{\lambda}(b)&=\lambda_{i}+\lambda_{j}'-i-j+1. 
\end{align}
For example, the Young diagram $\lambda$ $=$ $(5,4,2,2,1)$ for $N=14$ has the hook-lengths
\begin{align}
\label{hook_EG}
\young(97431,7521,42,31,1)
\end{align}

%%%%%%%%%%%%%%%%%%%%%%%%%%%%%%%%%%%%%%%%%%%
\subsubsection{Comparison with Schur functions}
\label{sec_ADHMN_C}
%%%%%%%%%%%%%%%%%%%%%%%%%%%%%%%%%%%%%%%%%%%
%Coulomb
The quantum Coulomb branch algebra $\mathcal{A}_{\epsilon_{1},\epsilon_{2}}^{C}$ of ADHM theory with one flavour 
is constructed from variables $w_{k,a}$, $v_{k,a}$ and has the monopole operators \cite{Kodera:2016faj}
\begin{align}
\label{mono_Ekt}
E_{k,t}&=\sum_{a=0}^{N_{k}}
\frac{\prod_{b} w_{k,a}-w_{k-1,b}-\epsilon_{2}}
{\prod_{b\neq a}w_{k,a}-w_{k,b}}w_{k,a}^{t}v_{k,a},
\\
\label{mono_Fkt}
F_{k,t}&=\sum_{a=0}^{N_{k}}
\frac{\prod_{b} w_{k,a}-w_{k+1,b}+\epsilon_{2}}{\prod_{b\neq a}w_{k,a}-w_{k,b}}
v_{k,a}^{-1}w_{k,a}^{t+\delta_{0,k}},
\end{align}
for a decomposition $\sum_{k}N_{k}=N$. 
The Verma module $V_{\lambda}$ of the quantum Coulomb branch algebra $\mathcal{A}^{C}_{\epsilon_{1},\epsilon_{2}}$ of the ADHM theory with one flavour is labelled by the Young diagram 
$\lambda = (\lambda_{1},\lambda_{2},\cdots)$. 
It is generated from vectors annihilated by all $F_{k,t}$ 
and has a basis of the form $|n_{k,a}\rangle$ which are eigenvectors for $w_{k,a}$ \cite{Gaiotto:2019wcc}
\begin{align}
\label{qCverma_w}
w_{k,a}|n_{k,a}\rangle&=
\left[
\epsilon_{2}c 
-(\epsilon_{1}+\epsilon_{2})r
-\epsilon_{1}n_{k,a}
\right]
|n_{k,a}\rangle
\end{align}
where $c$ and $r$ stand for the column and row to which the corresponding box belongs. 

We find that the twisted character of the Verma module $V_{\lambda}$ of the Coulomb branch algebra $\mathcal{A}_{\epsilon_{1},\epsilon_{2}}^{C}$ 
of the ADHM theory with one flavour is given by 
\begin{align}
\label{qC_trace_G}
\chi^{C}_{\epsilon_{1},\epsilon_{2},\lambda}(\beta)
&={\Tr}_{V_{\lambda}} (-1)^{\sum n_{k,a}}e^{-2\pi\beta \sum w_{k,a}}\nonumber\\
&=\frac{e^{2\pi \beta \epsilon_{1}n(\lambda)} e^{-2\pi \beta \epsilon_{2}\sum_{j}jn_{j}} }
{\prod_{b\in \lambda} \left(1-(-e^{2\pi \beta \epsilon_{1}})^{h_{\lambda}(b)}\right)}
\end{align}
where
\begin{align}
\label{nlambda}
n(\lambda)&=\sum_{i}(i-1)\lambda_{i}. 
\end{align}

For example, for $N=4$ with the decomposition $N_{0}=$ $N_{-1}=$ $N_{1}=$ $N_{2}=1$ we have the eigenvectors with
\begin{align}
\label{N4_eigenB}
w_{0,0}|n_{0,0}\rangle&=-\epsilon_{1}n_{0,0}|n_{0,0}\rangle,\nonumber\\
w_{-1,0}|n_{-1,0}\rangle&=
\left( -(\epsilon_{1}+\epsilon_{2})-\epsilon_{1}n_{-1,0} \right)|n_{-1,0}\rangle,\nonumber\\
w_{1,0}|n_{1,0}\rangle&=(\epsilon_{2}-\epsilon_{1}n_{1,0})|n_{1,0}\rangle,\nonumber\\
w_{2,0}|n_{2,0}\rangle&=(2\epsilon_{2}-\epsilon_{1}n_{2,0})|n_{2,0}\rangle. 
\end{align}
This Verma module is labelled by the Young diagram $\tiny \yng(3,1)$. 
We can directly compute the twisted Verma character as
\begin{align}
\label{qC_traceN4B}
&\chi^{C}_{\epsilon_{1},\epsilon_{2},\tiny \yng(3,1)}(\beta)
={\Tr}_{V_{\tiny \yng(3,1)}} (-1)^{n_{0,0}+n_{-1,0}+n_{1,0}+n_{2,0}} e^{-2\pi \beta (w_{0,0}+w_{-1,0}+w_{1,0}+w_{2,0})}
\nonumber\\
&=
\sum_{n_{0,0}=0}^{\infty}
\sum_{n_{-1,0}\ge n_{0,0}}
\sum_{n_{1,0}\ge n_{0,0}}
\sum_{n_{2,0}\ge n_{1,0}}
(-1)^{n_{0,0}+n_{-1,0}+n_{1,0}+n_{2,0}}
e^{2\pi\beta (\epsilon_{1}-2\epsilon_{2})}
e^{2\pi \beta \epsilon_{1}(n_{0,0}+n_{-1,0}+n_{1,0}+n_{2,0})}
\nonumber\\
&=
\frac{
e^{-4\pi \beta \epsilon_{2}+2\pi \beta \epsilon_{1}}
}{
(1+e^{2\pi \beta \epsilon_{1}})^2
(1-e^{4\pi \beta  \epsilon_{1}}) 
(1-e^{8\pi \beta  \epsilon_{1}}) 
}, 
\end{align}
which can be reproduced from the character formula (\ref{qC_trace_G}). 
The twisted Verma character (\ref{qC_traceN4B}) shows up in the residues (\ref{ADHM4_2}) 
at poles labelled by Young diagram $\tiny \yng(3,1)$ for $N=4$. 

As we expect, we see that the $S^3$ partition function (\ref{ADHMN_sum2}) takes the form 
of a sum over products of the twisted character (\ref{qC_trace_G}) of the Verma modules $V_{\lambda}$ 
for the quantum Higgs and Coulomb branch algebra, which are both identified with 
the spherical part of the rational Cherednik algebra associated with the Weyl group \cite{Kodera:2016faj}. 

Note that the Schur function of variables $(1,q,q^2,\cdots)$ can be expressed as \cite{MR325407}
\begin{align}
\label{schur_q}
s_{\lambda}(1,q,q^2,\cdots)
&=\frac{q^{n(\lambda)}}{\prod_{b\in \lambda}(1-q^{h_{\lambda}(b)})}
\end{align}
For example, for the Young diagram $\lambda$ $=$ $(5,4,2,2,1)$ for $N=14$ with the hook-lengths (\ref{hook_EG})
the Schur function (\ref{schur_q}) is
\begin{align}
\label{schur_qEG}
s_{\tiny (5,4,2,2,1)}(1,q,q^2,\cdots)
&=\frac{q^{18}}{(1-q)^{4} (1-q^{2})^{2} (1-q^3)^2 (1-q^4)^2 (1-q^5) (1-q^7)^2 (1-q^9)}. 
\end{align}
Making use of the formula (\ref{schur_q}), the Verma character (\ref{qC_trace_G}) can be expressed 
in terms of Schur function: 
\begin{align}
\label{qC_trace_G2}
\chi^{C}_{\epsilon_{1},\epsilon_{2},\lambda}(\beta)&=
(-1)^{n(\lambda)} e^{-2\pi \beta \epsilon_{2}\sum_{j}jn_{j}} 
s_{\lambda}\left(1,(-e^{2\pi \beta\epsilon_{1}}),(-e^{2\pi \beta\epsilon_{1}})^2,\cdots \right). 
\end{align}

%%%%%%%%%%%%%%%%%%%%%%%%%%%%%%%%%%%%%%%%%%%
\subsubsection{Reverse plane partition}
\label{sec_RPP}
%%%%%%%%%%%%%%%%%%%%%%%%%%%%%%%%%%%%%%%%%%%
%RPP
The expression (\ref{ADHMN_sum2}) provides us with an interesting view of the $S^3$ partition function of ADHM gauge theory in terms of a weak reverse plane partition. 
A reverse plane partition of a Young diagram $\lambda$ is a plane partition 
which fills the boxes with the non-negative integers in such a way that the entries in rows and columns are weakly increasing. 
When it admits $0$ as a part, it is called weak reverse plane partition. 

Let $\pi$ be weak reverse partition and $|\pi|$ be the sum of the entries in $\pi$. 
The generating function for weak reverse plane partitions of Young diagram $\lambda$ is given by \cite{MR325407} 
\footnote{
The generating function of all plane partitions is given by McMahon function
\begin{align}
\label{McMahon}
\sum_{\pi\in \mathrm{PP}} q^{|\pi|}&=\prod_{i\ge1}\frac{1}{(1-q^i)^i}. 
\end{align}
}
\begin{align}
\label{RPP_fcn}
\sum_{\pi\in \mathrm{RPP}} q^{|\pi|}&=\prod_{b\in \lambda} \frac{1}{1-q^{h_{\lambda}(b)}}. 
\end{align}

\begin{figure}
\begin{center}
\includegraphics[width=12.5cm]{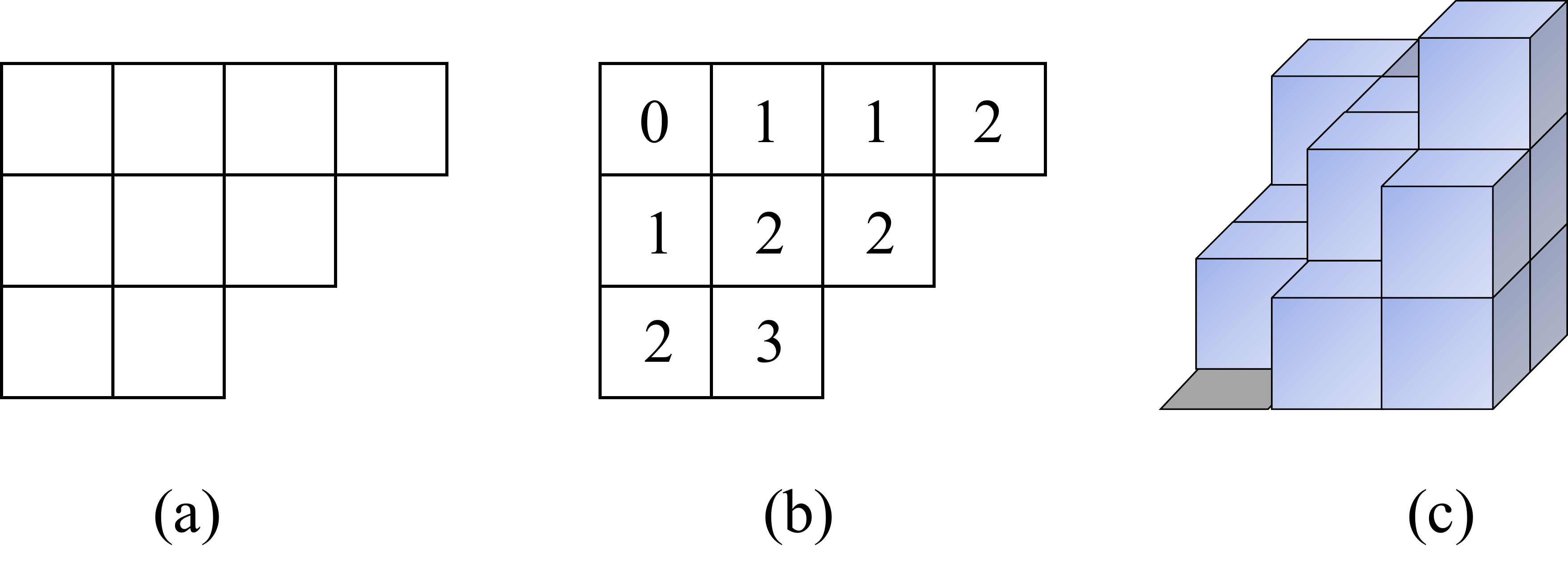}
\caption{
(a) Young diagram $\lambda=(4,3,2)$. 
(b) Reverse plane partition $\pi$. 
(c) Crystal built from $\pi$ with $|\pi|=14$ blocks. 
}
\label{figRPP}
\end{center}
\end{figure}

We can view the numbers in the entries in weak reverse partition as the heights of blocks placed on each box of the Young diagram $\lambda$. 
Then we can build up the associated three-dimensional crystal for $\pi$ 
where the total number of blocks is equal to $|\pi|$ (see Figure \ref{figRPP}). 

In terms of the generating function of weak reverse plane partitions, 
we can express the Verma character (\ref{qC_trace_G}) of Coulomb branch algebra $\mathcal{C}_{\epsilon_{1},\epsilon_{2}}^{(N)}$ as
\begin{align}
\label{qC_trace_G3}
\chi^{C}_{\epsilon_{1},\epsilon_{2},\lambda}(\beta)
&=
e^{-2\pi \beta\epsilon_{2}\sum_{j} j n_{j}}
e^{2\pi \beta \epsilon_{1}n(\lambda)} 
\sum_{\pi\in \mathrm{RPP}}(-e^{2\pi \beta \epsilon_{1}|\pi|})
\end{align}
and the $S^3$ partition function (\ref{ADHMN_sum2}) of ADHM theory as
\begin{align}
\label{ADHMN_sum3}
Z_{\mathrm{ADHM}^N} &= 
\sum_{\lambda}
\frac{C_{1}}{C_{2}C_{3}} 
e^{\pi (m+\zeta)\sum_{k=1}^{\infty} k d_{k}}
e^{-2\pi i\zeta m \sum_{j}jn_{j}}
\sum_{\pi\in \mathrm{RPP}} (-e^{2\pi m |\pi|}) 
\sum_{\pi\in \mathrm{RPP}} (-e^{2\pi \zeta |\pi|}). 
\end{align}

%%%%%%%%%%%%%%%%%%%%%%%%%%%%%%%%%%%
\subsection*{Acknowledgements}
We would like to thank 
Mykola Dedushenko, Tudor Dimofte, Jaume Gomis, Kazuo Hosomichi, Joel Kamnitzer, Michael McBreen, Silviu Pufu, Leonardo Rastelli and Junya Yagi 
for useful discussions and comments. 
D.G. is supported by the Perimeter Institute for Theoretical Physics.
T.O. is supported in part by Perimeter Institute for Theoretical Physics and 
JSPS Overseas Research fellowships. Research at
Perimeter Institute is supported by the Government of Canada through the Department of
Innovation, Science and Economic Development and by the Province of Ontario through the
Ministry of Research, Innovation and Science.
%%%%%%

\appendix
\section{Comparison to the Schur correlators} \label{app:sh}
The Schur index of a 4d theory is an elliptic variant of the $S^3$ partition function. The $2\cosh \pi x$ factors are replaced by 
theta functions 
\begin{equation}
\prod_{i=0}^\infty (1+\xi q^{n + \frac12})(1+\xi^{-1} q^{n + \frac12})
\end{equation}
and the Vandermonde $2\sinh 2\pi x$ by 
\begin{equation}
(\xi - \xi^{-1}) \prod_{i=1}^\infty (1-\xi q^n)(1-\xi^{-1} q^{n}). 
\end{equation}
The contour integral runs along $|\xi| = 1$ contours, as it is a projection on gauge invariant operators. The measure is 
\begin{equation}
 \prod_{i=1}^\infty (1- q^n)\frac{d\xi}{2 \pi i \xi}. 
\end{equation}
FI parameters are usually not included, as $U(1)$ gauge factors are IR free and have Landau poles. They would insert some $\xi^\zeta$ factor which 
would be anyway troublesome with the standard integration contour unless $\zeta$ is integer. 

The Coulomb branch (line) operators admit an Abelianized description analogous to the one for the 3d quantized Coulomb branch, 
\begin{equation}
L_C = \sum_{n_*} R_{n_*}(\xi_*,\mu) v_{n_*}
\end{equation}
except that the $v_{n_*}$ multiply $\xi_i$ by $q^{n_i}$. Again, the Schur correlation function is computed by inserting $R_{0}(\xi_*,\mu)$ in the contour integral.

The twisted trace property still holds, but the twisting involves a rotation by $U(1)_r$ by $2 \pi$. When the theory is not conformal, $U(1)_r$
itself is anomalous but the twisting transformation is still well-defined. In gauge theories, for example, it maps to some integral shift of the $\theta$ angle, 
which shifts the electric charges of monopole line defects. 

For theories of class ${\cal S}$ associated to a surface $\Sigma$ and Lie algebra $\mathfrak{g}$, the quantized Coulomb branch algebra is the quantization of the algebra of functions on the character variety
of complex flat $\mathfrak{g}$ local systems on $\Sigma$.

Concretely, generators can be depicted as closed networks of $\mathfrak{g}$ Wilson lines drawn in $\Sigma \times \mathbb{R}$, modulo skein relations for $\mathfrak{g}$. They are composed simply by 
concatenation along the $\mathbb{R}$ direction. A natural way to produce traces is to consider some kind of (analytically continued) Chern-Simons theory on $\Sigma \times S^1$,
adjusted to account for the twisting. It should be possible to connect such a construction to the Schur index, perhaps using the strategy of \cite{Cordova:2013cea,Mikhaylov:2017ngi}. 

\section{Specializing $S^3_b$}\label{app:Sb}
A 3d ${\cal N}=2$ chiral multiplet of mass $x$ contributes a factor of \cite{Hama:2011ea}
\begin{equation}
s_b(\frac{i b}{2}+\frac{i}{2b}- x) \sim \frac{\prod_{m,n=1}^\infty (m b + n b^{-1}+ i x)}{\prod_{m,n=0}^\infty (m b + n b^{-1}- i x)}
\end{equation}
to the ellipsoid integral.

An hypermultiplet will contribute 
\begin{equation}
s_b(\frac{i b}{2}+\frac{i}{2b}- x-y)s_b(\frac{i b}{2}+\frac{i}{2b}+ x-y) = \frac{s_b(\frac{i b}{2}+\frac{i}{2b}+ x-y) }{s_b(x+y-\frac{i b}{2}-\frac{i}{2b}) }
\end{equation}, where $y$ is the mass for the 
extra R-symmetry generator.

If we set $y = \frac{i}{2b}$, this reduces to 
\begin{equation}
\frac{s_b(\frac{i b}{2}+ x) }{s_b(x-\frac{i b}{2}) } = \frac{1}{2 \cosh \pi b x}
\end{equation}
which is what appears in the special sphere partition function, with $b x \to x$. 

In a similar way, a  3d ${\cal N}=2$ gauge multiplet contributes a Vandermonde factor of $4 \sinh \pi b x \sinh \pi b^{-1} x$. 
The full vectormultiplet corrects that to 
\begin{equation}
s_b(2 y - \frac{i b}{2}- \frac{i}{2b}+ x)s_b(2 y - \frac{i b}{2}- \frac{i}{2b}-x) \sinh \pi b x \sinh \pi b^{-1} x
\end{equation}
which simplifies at $y = \frac{i}{2b}$ to the desired $4 \sinh^2 \pi b x$. 

\bibliographystyle{utphys}
\bibliography{ref}

%\bibliographystyle{JHEP}
%\bibliography{mono}

\end{document}